\documentclass[12pt,a4paper]{article}
\usepackage{a4wide,latexsym,graphicx,epsfig,psfrag,float}
\usepackage{amsmath}
\usepackage{longtable}
\usepackage{amssymb}
\usepackage{bbm}
\usepackage{hyperref}
\allowdisplaybreaks

\usepackage{color}

\newcommand{\lsim}{\stackrel{<}{_\sim}}

\providecommand{\openone}{\leavevmode\hbox{\small1\kern-3.8pt\normalsize1}}

\begin{document}
\parskip=3pt plus 1pt

\begin{titlepage}
\begin{flushright}
{
JLAB-THY-13-1733 \\
IFJPAN-IV-2013-3 \\ 
FTUV/13-0524 \\
IFIC/13-19
}
\end{flushright}
\vskip 2cm

\setcounter{footnote}{0}
\renewcommand{\thefootnote}{\fnsymbol{footnote}}

\begin{center}
{\LARGE \bf Three pseudoscalar meson production}
\\ [13pt]{\LARGE \bf in  $e^+ e^-$ annihilation}
\vspace{2.1cm} \\
{\sc  L.Y.~Dai}$^{a,b}$\footnote{Email:~lingyun@jlab.org}{\sc , J.~Portol\'es}$^{c}$\footnote{Email:~Jorge.Portoles@ific.uv.es} and
{\sc O.~Shekhovtsova}$^{d,e}$\footnote{Email:~olga.shekhovtsova@ifj.edu.pl}
\vspace{1.3cm} \\
$^{a}$ Department of Physics, Peking University, Beijing 1000871, China \\ [7pt]
$^{b}$ Thomas Jefferson National Accelerator Facility, Newport News, Virginia 23606, USA \\ [7pt]
$^{c}$ IFIC, CSIC - Universitat de Val\`encia, Apt. Correus 22085, E-46071 Val\`encia, Spain \\ [7pt]
$^{d}$ NSC Kharkov Institute of Physics and Technology, Kharkov UA-61108, Ukraine \\ [7pt]
$^{e}$ Institute of Nuclear Physics, PAN, Krakow, ul. Radzikowskiego 152, Krakow, Poland
\end{center}

\setcounter{footnote}{0}
\renewcommand{\thefootnote}{\arabic{footnote}}
\vspace*{0.9cm}

\begin{abstract}
We study, at leading order in the large number of colours expansion and within the Resonance Chiral Theory framework,
the odd-intrinsic-parity $e^+ e^- \rightarrow \pi^+ \pi^- (\pi^0 , \eta)$
cross-sections in the energy regime populated by hadron resonances, namely $3 \, m_{\pi} \lsim E \lsim  2 \, \mbox{GeV}$.
In addition we implement our results in the Monte Carlo generator PHOKHARA 7.0 and we simulate hadron production through the radiative return method.
\end{abstract}
\vfill
PACS~: 11.30.Rd, 12.39.Fe, 13.66.Bc \\
Keywords~: Chiral symmetry, Effective Lagrangian, Large-$N_C$.

\end{titlepage}

\section{Introduction}
\label{sec:1}
The hadronization of currents in the non-perturbative regime of QCD provides important information on the dynamics of hadrons. Hadron currents associated with physical
processes are parameterized in terms of functions of the kinematical invariants of the amplitude, namely the form factors. Though the dynamics involved is that of the strong interaction
the fact that QCD is not perturbative below, let us say, $2 \, \mbox{GeV}$ implies that their determination is limited by our lack of knowledge of QCD in that
regime. Nevertheless at very low energies, $E \ll M_{\rho}$ ($M_\rho$ being the mass of the lightest hadron resonance, $\rho(770)$), Chiral Perturbation Theory ($\chi$PT)
\cite{Weinberg:1978kz,Gasser:1983yg}, based on the chiral symmetry of massless QCD, provides a rigorous Effective Field Theory (EFT) framework that 
strongly constrains the structure of the above mentioned form factors.
\par
 At higher energies, $M_{\rho} \lsim E \lsim 2 \, \mbox{GeV}$, we lack such a framework and so have to rely on phenomenological
approaches. Resonance Chiral Theory (R$\chi$T) \cite{Ecker:1988te,Ecker:1989yg,Cirigliano:2006hb,Portoles:2010yt} provides a useful tool to handle the study of
hadronization procedures in the energy region dominated by resonances. It relies upon a phenomenological Lagrangian system that includes the relevant degrees of
freedom (pseudo-Goldstone mesons and resonances) and the Large-$N_C$ expansion \cite{'tHooft:1973jz,'tHooft:1974hx,Witten:1979kh} that defines the scheme. This framework
is known to provide a successful approach to the dynamics driven by the lightest hadron resonances \cite{RuizFemenia:2003hm,Cirigliano:2004ue,GomezDumm:2003ku,Cirigliano:2005xn}.
Indeed it has been applied to the study of hadron tau decays \cite{GomezDumm:2003ku,Dumm:2009va,Dumm:2009kj} -where results have recently been implemented to upgrade the Monte Carlo generator TAUOLA \cite{Shekhovtsova:2012ra}-, two-photon transition form factors \cite{Czyz:2012nq} and their implementation into the code EKHARA, and to the hadronization of the vector current in
$e^+ e^-$ annihilation \cite{Dubinsky:2004xv}. This latter case is our interest here. Though the $e^+ e^- \rightarrow PP$ scattering process ($P$ is short for a pseudoscalar meson) has
been widely analyzed through different approaches to the vector form factor both at low energies \cite{Gasser:1984ux,Bijnens:1998fm} and in the resonance energy region
\cite{Guerrero:1997ku,SanzCillero:2002bs,Pich:2001pj}, the study of more than two mesons in the final
state is not so well developed. The case of three pions has been approached in Ref.~\cite{Czyz:2005as} through a phenomenological analysis using vector meson dominance.
A thorough study of the production of four pions within the R$\chi$T framework has been performed \cite{Ecker:2002cw}, while an analysis of this channel based on a phenomenological Lagrangian  parameterization has also been provided \cite{Czyz:2000wh,Czyz:2008kw}. Exclusive channels with higher number of mesons in the final state have not been considered, though a  non-dynamical
approach based on isospin symmetry could be applied \cite{Sobie:1995ba,Sobie:1999ke} as a first strategy.
\par
The experimental measurement of the $e^+ e^- \rightarrow \pi^+ \pi^- \pi^0$ cross-section has been carried out some time ago \cite{Cordier:1979qg,Antonelli:1992jx}
(see \cite{Whalley:2003qr} for a complete list of references up to 2003). Recently there has been new interest in this process by the SND  \cite{Achasov:2003ir} and CMD-2
\cite{Akhmetshin:2003zn} Collaborations. Finally, BaBar has provided a measurement of this cross-section \cite{Aubert:2004kj} by considering the initial state radiation (ISR)
\cite{Arbuzov:1998te,Binner:1999bt,Benayoun:1999hm}. The $\pi^+ \pi^- \eta$ final state has a considerably shorter history and only DM2 \cite{Antonelli:1988fw}
and BaBar \cite{Aubert:2007ef} have collected data over all the energy region populated by resonances.
\par
The analysis of $e^+ e^- \rightarrow \pi^+ \pi^- \pi^0$ carried out in the framework of vector meson dominance in Ref.~\cite{Czyz:2005as} did not include the anomalous structure of
the production of three mesons from a vector current, as given by QCD. This vertex violates intrinsic-parity conservation and is driven, at very low energies, by the QCD anomaly that
contributes at the leading chiral ${\cal O}(p^4)$. Starting at ${\cal O}(p^6)$ and well into the resonance energy region (where the chiral counting no longer
applies), the dynamics is driven by non-anomalous contributions.  In this article we consider both contributions and our result is then consistent with our present knowledge of QCD.
\par
In Section~\ref{sec:2} we specify our theoretical framework. The analyses of the form factors that determine the cross-sections of interest are performed
in Section~\ref{sec:3}. Then we improve our approach by taking into account short-distance constraints on the unknown couplings of the phenomenological
Lagrangian, Section~\ref{sec:4}, and including corrections such as isospin breaking and the total widths of resonances that improve the original form factors in
Section~\ref{sec:5}.
We comment on the description of the partial decay widths that involve vector resonances and that have been taken into account in the global analysis in Section~\ref{sec:7}. The fit procedure and results are detailed in Section~\ref{sec:6}. In Section~\ref{sec:8} we
explain the implementation of our results in PHOKHARA 7.0. Our final conclusions are collected in Section~\ref{sec:9}. Finally two appendices
gather most of the relevant analytical results of our work.

\section{Theoretical framework: Resonance Chiral Theory}
\label{sec:2}
Chiral Perturbation Theory  provides a rigorous EFT framework for long-distance QCD in the energy region where resonances still have not yet
arisen ($E \ll M_{\rho}$). In the energy region
where resonances live we do not have such an EFT framework, but we can approach the chiral setting by a phenomenological Lagrangian where both Goldstone modes and the fields describing
the meson resonances are active degrees of freedom. R$\chi$T \cite{Ecker:1988te,Ecker:1989yg,Cirigliano:2006hb} provides such a framework and its main features can be summarized like this: i) It contains $\chi$PT, hence the amplitudes of R$\chi$T satisfy chiral symmetry and match the chiral amplitudes at low energies; ii) The high energy behaviour of form factors provided by
R$\chi$T matches their known smearing and suppression enforced by QCD. Therefore we work in a landscape that matches appropriately both the better
 known low and high energy domains, and
that we know how to describe. We follow closely the notation of Ref.~\cite{Cirigliano:2006hb} where definitions not stated here can be found.
\par
Massless QCD, a suitable starting point for the three lightest flavours: $u$, $d$ and $s$, is invariant under $G = SU(3)_L \otimes SU(3)_R$ global transformations on the left- and
right-handed quarks, separately, in flavour space. In addition the chiral group $G$ is spontaneously broken to the diagonal subgroup $SU(3)_V$ and the eight pseudoscalar massless bosons,
to be identified with the lightest mesons, arise. The Goldstone fields $\Phi$ parameterize the elements $u(\Phi)$ of the coset space $SU(3)_L \otimes SU(3)_R / SU(3)_V$. An
explicit parameterization of these elements is given by:
\begin{eqnarray} \label{eq:uphi}
 u(\Phi) = \exp \left\{ \frac{i}{\sqrt{2} F} \Phi \right\} \, ,
\end{eqnarray}
with
$$
\Phi =  \frac{1}{\sqrt{2}} \sum_{i=1}^8 \lambda_i \varphi_i =
\left(
\begin{array}{ccc}
 \displaystyle\frac{1}{\sqrt 2}\,\pi^0 + \displaystyle\frac{1}{\sqrt
 6}\,\eta_8
& \pi^+ & K^+ \\
\pi^- & - \displaystyle\frac{1}{\sqrt 2}\,\pi^0 +
\displaystyle\frac{1}{\sqrt 6}\,\eta_8
& K^0 \\
 K^- & \bar{K}^0 & - \displaystyle\frac{2}{\sqrt 6}\,\eta_8
\end{array}
\right)
\ ,
$$
and $F \simeq F_{\pi} = 92.4 \, \mbox{MeV}$ the decay constant of the pion. In fact, as we are also interested in the study of the $\eta, \eta'$ production, we will consider
a $U(3)$ symmetry instead of $SU(3)$ by substituting $\Phi \rightarrow \Phi + \eta_0/\sqrt{3} \, \mathbbm{1}$. The mixing of the $\eta_8$ and $\eta_0$ to give the physical $\eta$ and $\eta'$ states is defined through the $\theta_P$ mixing angle by:
\begin{equation} \label{eq:etas}
 \left( \begin{array}{c}
          \eta_8 \\ \eta_0
        \end{array}
 \right)
= \left( \begin{array}{cc}
          \cos \theta_P & \sin \theta_P \\ -\sin \theta_P & \cos \theta_P
         \end{array}
 \right) \, \left( \begin{array}{c}
                    \eta \\ \eta'
                   \end{array}
\right).
\end{equation}
The introduction of the resonance fields follows from their consideration as matter fields \cite{Coleman:1969sm,Callan:1969sn}. The nonlinear realization of G
on these depends on their transformation properties under the unbroken subgroup $SU(3)_V$. We consider states transforming as octets and singlets:
\begin{equation} \label{eq:transos}
R_{\bf{8}} \rightarrow h(g,\Phi) \,   R_{\bf{8}}  \, h(g,\Phi)^{-1} \ , \qquad \qquad
\qquad R_{\bf{0}} \, \rightarrow \,    R_{\bf{0}} \ ,
\end{equation}
in terms of the $SU(3)_V$ compensator field $h(g,\Phi)$, $g=(g_L,g_R) \in G$.
In the large-$N_C$
limit, both octet and singlet become degenerate in the chiral limit and we collect them in a nonet field,
\begin{equation} \label{eq:rnonet}
 R  = \sum_{i=1}^8 \frac{\lambda_i}{\sqrt{2}} \,  R_i + \frac{R_{\bf{0}}}{\sqrt{3}} \, \mathbbm{1} \, .
\end{equation}
Particularizing to vector meson resonances, the relation between the octet and nonet components and the physical $\omega(782)$ and $\phi(1020)$ states is given by
the $\theta_V$ mixing angle through:
\begin{equation} \label{eq:vectos}
 \left( \begin{array}{c}
          V^8_{\mu} \\ V^0_{\mu}
        \end{array}
 \right)
= \left( \begin{array}{cc}
          \cos \theta_V & \sin \theta_V \\ -\sin \theta_V & \cos \theta_V
         \end{array}
 \right) \, \left( \begin{array}{c}
                    \phi_{\mu} \\ \omega_{\mu}
                   \end{array}
\right).
\end{equation}
\par
Finally in order to be able to provide Green functions of the different quark currents, it is convenient to include external hermitian fields coupled to them that transform
locally under the G group, namely vector $v_{\mu}(x)$, axial-vector $a_{\mu}(x)$, scalar $s(x)$ and pseudoscalar $p(x)$ sources. Once we have this field and external sources content, together
with their transformation properties, the hadronic Lagrangian proceeds through the inclusion of the most general set of operators
invariant under Lorentz, chiral, P and C transformations:
\begin{equation} \label{eq:rtypo}
 {\cal O}_i \sim  \langle R_a...R_f \chi^{(n)}(\Phi) \rangle \, ,
\end{equation}
where the brackets $\langle \rangle$ indicate a trace in the flavour space and
$\chi^{(n)}(\Phi)$ is a chiral tensor that includes only the Goldstone and external fields and that transforms like $R_{\bf{8}}$ in Eq.~(\ref{eq:transos}) or remain invariant
under chiral transformations. $\chi^{(n)}(\Phi)$ is of ${\cal O}(p^n)$ in the chiral counting and its building blocks: $u_{\mu}$, $\chi_{\pm}$, $f_{\pm}^{\mu \nu}$ and
$h_{\mu \nu}$, can be read from Refs.~\cite{Ecker:1988te,Ecker:1989yg,Cirigliano:2006hb}.
\par
As we will comment later on, our study of the hadronization of the vector current in the production of three mesons in $e^+ e^-$ collisions, will be driven by the
$N_C \rightarrow \infty$ limit of large-$N_C$ QCD. In the construction of our Lagrangian we will follow the canonical counting: terms with single traces are of order
$N_C$, while additional traces reduce the order in $1/N_C$ each. For a discussion of this point, see Appendix A of Ref.\cite{Cirigliano:2006hb}. Hence our construction
of the operators ${\cal O}_i$ in (\ref{eq:rtypo}) will involve only one flavour trace.
\par
We specify now our R$\chi$T Lagrangian. It consists of pieces that only involve pseudoscalar Goldstone bosons: ${\cal L}^{\mbox{\tiny GB}}$, and those that include their couplings to resonances: ${\cal L}_{\mbox{\tiny R}}$.
The first one contains the leading  ${\cal O}(p^ 2)$ $\chi$PT Lagrangian and higher order operators with the same structure that those
of $\chi$PT. They are identified by their chiral order: ${\cal L}^{\mbox{\tiny GB}} = {\cal L}^{\mbox{\tiny GB}}_{(2)} + {\cal L}^{\mbox{\tiny GB}}_{(4)} + \ldots$. As was shown in Ref.~\cite{Ecker:1989yg}, the use of antisymmetric
tensors to describe spin-1 resonances in ${\cal L}_{\mbox{\tiny R}}$ provides a QCD consistent description of two-point Green functions and two-body form factors without the need to include any operator of dimension 4, i.e. ${\cal L}^{\mbox{\tiny GB}}_{(4)}=0$ in the even-intrinsic-parity sector. In other words, the low-energy constants of the
${\cal O}(p^4)$ Goldstone boson Lagrangian are saturated by the resonance couplings if one uses the antisymmetric description of the spin-1 resonances. Hence we will use that formulation in our setting.
Accordingly our Goldstone boson Lagrangian consists in the ${\cal O}(p^2)$ $\chi$PT Lagrangian:
\begin{equation} \label{eq:chpt2}
 {\cal L}^{\mbox{\tiny GB}}_{(2)} \equiv {\cal L}^{\mbox{\tiny $\chi$PT}}_{(2)} =  \frac{F^2}{4} \langle u_{\mu} u^{\mu} + \chi_+ \rangle \, ,
\end{equation}
and the ${\cal O}(p^4)$ chiral anomaly functional $Z[U, v, a]$ \cite{Witten:1983tw}, with $U= u^2$, that implements the Wess-Zumino anomaly \cite{Wess:1971yu}. This is the
leading odd-intrinsic-parity contribution and we collect here the piece that is relevant for our goal:
\begin{equation} \label{eq:wzanomaly}
 {\cal L}^{\mbox{\tiny GB}}_{(4)} = i \frac{N_C \sqrt{2}}{12 \pi^2 F^3} \, \varepsilon_{\mu \nu \rho \sigma} \, \langle \partial^{\mu} \Phi \, \partial^{\nu} \Phi \, \partial^{\rho} \Phi \,  v^\sigma \rangle + ... \, .
\end{equation}
Notice that the pure Goldstone boson Lagrangian is completely specified by the decay constant of the pion, the number of colours $N_C$ and the masses of the lightest pseudoscalar
mesons. The latter are encoded into the scalar external field $s(x) = {\cal M} + ...$ in the $\chi_{+}$ chiral tensor, being ${\cal M}=\mbox{diag}(m_u,m_d,m_s)$ the diagonal quark mass
matrix.
\par
Let us add now the part of the R$\chi$T Lagrangian that includes the resonance fields and their coupling to the lightest pseudoscalar mesons. We will only need to include, for the study
of our processes, pieces with vector resonances:
\begin{equation} \label{eq:lr}
 {\cal L}_{\mbox{\tiny V}} = {\cal L}^{\mbox{\tiny V}}_{\mbox{\tiny kin}} + {\cal L}^{\mbox{\tiny V}}_{\mbox{\tiny int}} \, .
\end{equation}
Here:
\begin{equation}\label{eq:kin}
 {\cal L}^{\mbox{\tiny V}}_{\mbox{\tiny kin}} = - \frac{1}{2} \langle \nabla^{\lambda} V_{\lambda \mu} \nabla_{\nu} V^{\nu \mu} \rangle + \frac{1}{4} M_V^2 \langle
V_{\mu \nu} V^{\mu \nu} \rangle  \, ,
\end{equation}
with the covariant derivative $\nabla_{\mu}$ defined in Ref.~\cite{Ecker:1988te}
and ${\cal L}^{\mbox{\tiny V}}_{\mbox{\tiny int}} = {\cal L}^{\mbox{\tiny V}}_{(2)}+{\cal L}^{\mbox{\tiny V}}_{(4)}+{\cal L}^{\mbox{\tiny VV}}_{(2)}$, where the operators are like
those in Eq.~(\ref{eq:rtypo}), $n$ in ${\cal L}_{(n)}$ indicates the order of the chiral tensor and the super index indicates the number of vector mesons. Operators with chiral tensor $n=2$, linear
in the vector fields, are of the even-intrinsic-parity type \cite{Ecker:1988te}:
\begin{equation} \label{eq:lv2}
{\cal L}^{\mbox{\tiny V}}_{(2)} = \frac{F_V}{2 \sqrt{2}} \, \langle V_{\mu \nu} f_+^{\mu \nu} \rangle + i \frac{G_V}{\sqrt{2}} \, \langle V_{\mu \nu} u^{\mu} u^{\nu} \rangle \, ,
\end{equation}
while the rest of operators, that contribute to the processes of our interest, are of odd-intrinsic-parity:
\begin{eqnarray} \label{eq:lv4}
 {\cal L}^{\mbox{\tiny V}}_{(4)} = \sum_{j=1}^7 \frac{c_j}{M_V} \, {\cal O}_{\mbox{\tiny VJP}}^j \, + \, \sum_{j=1}^5 \frac{g_j}{M_V} \, {\cal O}_{\mbox{\tiny VPPP}}^j \, ,
\end{eqnarray}
with \cite{RuizFemenia:2003hm,Dumm:2009kj}:
\begin{eqnarray}\label{eq:VJP}
{\cal O}_{\mbox{\tiny VJP}}^1 & = & \varepsilon_{\mu\nu\rho\sigma}\,
\langle \, \{V^{\mu\nu},f_{+}^{\rho\alpha}\} \nabla_{\alpha}u^{\sigma}\,\rangle
\; \; , \nonumber\\[3mm]
{\cal O}_{\mbox{\tiny VJP}}^2 & = & \varepsilon_{\mu\nu\rho\sigma}\,
\langle \, \{V^{\mu\alpha},f_{+}^{\rho\sigma}\} \nabla_{\alpha}u^{\nu}\,\rangle
\; \; , \nonumber\\[3mm]
{\cal O}_{\mbox{\tiny VJP}}^3 & = & i\,\varepsilon_{\mu\nu\rho\sigma}\,
\langle \, \{V^{\mu\nu},f_{+}^{\rho\sigma}\}\, \chi_{-}\,\rangle
\; \; , \nonumber\\[3mm]
{\cal O}_{\mbox{\tiny VJP}}^4 & = & i\,\varepsilon_{\mu\nu\rho\sigma}\,
\langle \, V^{\mu\nu}\,[\,f_{-}^{\rho\sigma}, \chi_{+}]\,\rangle
\; \; , \nonumber\\[3mm]
{\cal O}_{\mbox{\tiny VJP}}^5 & = & \varepsilon_{\mu\nu\rho\sigma}\,
\langle \, \{\nabla_{\alpha}V^{\mu\nu},f_{+}^{\rho\alpha}\} u^{\sigma}\,\rangle
\; \; ,\nonumber\\[3mm]
{\cal O}_{\mbox{\tiny VJP}}^6 & = & \varepsilon_{\mu\nu\rho\sigma}\,
\langle \, \{\nabla_{\alpha}V^{\mu\alpha},f_{+}^{\rho\sigma}\} u^{\nu}\,\rangle
\; \; , \nonumber\\[3mm]
{\cal O}_{\mbox{\tiny VJP}}^7 & = & \varepsilon_{\mu\nu\rho\sigma}\,
\langle \, \{\nabla^{\sigma}V^{\mu\nu},f_{+}^{\rho\alpha}\} u_{\alpha}\,\rangle
\;,
\end{eqnarray}
\begin{eqnarray}\label{eq:VPPP}
{\cal O}_{\mbox{\tiny VPPP}}^1 & = & i \, \varepsilon_{\mu\nu\alpha\beta} \, \langle
V^{\mu\nu} \, ( \, h^{\alpha\gamma} u_{\gamma} u^{\beta} - u^{\beta} u_{\gamma}
h^{\alpha\gamma} ) \rangle \, , \nonumber  \\ [2mm]
{\cal O}_{\mbox{\tiny VPPP}}^2 & = & i \, \varepsilon_{\mu\nu\alpha\beta} \, \langle
V^{\mu\nu} \, ( \, h^{\alpha\gamma} u^{\beta} u_{\gamma} - u_{\gamma} u^{\beta}
h^{\alpha\gamma} \, ) \rangle \, , \nonumber \\ [2mm]
{\cal O}_{\mbox{\tiny VPPP}}^3 & = & i \, \varepsilon_{\mu\nu\alpha\beta} \, \langle
V^{\mu\nu} \, ( \, u_{\gamma} h^{\alpha\gamma} u^{\beta}  -  u^{\beta}
h^{\alpha\gamma} u_{\gamma} \, ) \rangle \, , \nonumber \\ [2mm]
{\cal O}_{\mbox{\tiny VPPP}}^4 & = &  \varepsilon_{\mu\nu\alpha\beta} \, \langle
\lbrace  \,V^{\mu\nu} \,,\,  u^{\alpha}\, u^{\beta}\, \rbrace \,{\cal \chi}_{-} \rangle \, , \nonumber \\ [2mm]
{\cal O}_{\mbox{\tiny VPPP}}^5 & = &  \varepsilon_{\mu\nu\alpha\beta} \, \langle
 \,u^{\alpha}\,V^{\mu\nu} \, u^{\beta}\, {\cal \chi}_{-} \rangle \, .
\end{eqnarray}
Finally the terms quadratic in the vector field are \cite{RuizFemenia:2003hm}:
\begin{eqnarray} \label{eq:lvv2}
{\cal L}^{\mbox{\tiny VV}}_{(2)} = \sum_{j=1}^4 d_j {\cal O}_{\mbox{\tiny VVP}}^j,
\end{eqnarray}
\begin{eqnarray}\label{eq:VVP}
{\cal O}_{\mbox{\tiny VVP}}^1 & = & \varepsilon_{\mu\nu\rho\sigma}\,
\langle \, \{V^{\mu\nu},V^{\rho\alpha}\} \nabla_{\alpha}u^{\sigma}\,\rangle
\; \; , \nonumber\\[2mm]
{\cal O}_{\mbox{\tiny VVP}}^2 & = & i\,\varepsilon_{\mu\nu\rho\sigma}\,
\langle \, \{V^{\mu\nu},V^{\rho\sigma}\}\, \chi_{-}\,\rangle
\; \; , \nonumber\\[2mm]
{\cal O}_{\mbox{\tiny VVP}}^3 & = & \varepsilon_{\mu\nu\rho\sigma}\,
\langle \, \{\nabla_{\alpha}V^{\mu\nu},V^{\rho\alpha}\} u^{\sigma}\,\rangle
\; \; , \nonumber\\[2mm]
{\cal O}_{\mbox{\tiny VVP}}^4 & = & \varepsilon_{\mu\nu\rho\sigma}\,
\langle \, \{\nabla^{\sigma}V^{\mu\nu},V^{\rho\alpha}\} u_{\alpha}\,\rangle
\; .
\end{eqnarray}
Notice that, in contrast to the Goldstone Boson Lagrangian, the vector resonance operators introduce 17 unknown couplings
that are not given by symmetry assumptions alone. We will see in Section~\ref{sec:4} how it is possible to get information
on these couplings in order to minimize the number of parameters left for the fit to the experimental data. Though there
is nothing inherently wrong in leaving all the couplings free for the fit, we will see that there is information coming
from QCD that we can enforce on the Lagrangian couplings. This implementation would provide the free fit parameters with closer
values to those given by QCD, if we only knew how to get them.

\section{The amplitudes for $e^+ e^- \rightarrow \pi^+ \pi^- (\pi^0, \eta)$}
\label{sec:3}
The amplitude for three meson production in $e^+ e^-$ collisions at low energies is dominantly driven by the electromagnetic current ${\cal J}_{\mu}^{\rm{em}} =
{\cal V}_{\mu}^3+ {\cal V}_{\mu}^8/\sqrt{3}$ with ${\cal V}_{\mu}^i = \overline{q} \gamma_{\mu} \frac{\lambda^i}{2} q$, where $q$ represents the SU(3) multiplet of light quarks and
$\lambda^i$ are the Gell-Mann matrices.
The final states have definite $G$-parity, $G(\pi^+ \pi^- \pi^0) = -1$, $G(\pi^+ \pi^- \eta)= 1$ and  they
are given, in the isospin limit, by the isoscalar and isovector components, respectively.  However these contributions mix when isospin violating effects
are taken into account. The amplitude for the process
 $e^+(k') e^-(k) \rightarrow \pi^+(k_1) \pi^-(k_2) P(k_3)$ ($P= \pi^0, \eta$) is given by:
\begin{eqnarray} \label{eq:amplitude}
 {\cal M}^{P} &=&  - \frac{4 \pi \alpha}{Q^2}  \,  T_{\mu}^{P} \, \overline{v}(k') \gamma^{\mu} u(k)\, , \nonumber \\
T_{\mu}^{P} & \equiv & \langle  \,  \pi^+(k_1) \pi^-(k_2) P(k_3) \, |  {\cal J}_{\mu}^{\rm{em}} \, e^{i {\cal L}_{\mbox{\tiny QCD}}}| 0 \rangle ,
\end{eqnarray}
where $Q=k+k'$, being $E_{\mbox{\tiny CM }} \equiv \sqrt{Q^2}$ the energy in the center-of-mass system, and the $e^{i {\cal L}_{\mbox{\tiny QCD}}}$ factor reminds us that the matrix element, the hadronization of the electromagnetic current, has to be evaluated in the presence of strong interactions. Assuming parity symmetry, the hadronic matrix element is parameterized by one form factor only:
\begin{equation} \label{eq:ffdef}
  \langle \, \pi^+(k_1) \pi^-(k_2) P(k_3) \, |  {\cal J}_{\mu}^{\rm{em}} \, e^{i {\cal L}_{\mbox{\tiny QCD}}}| 0 \rangle \, = \, i F_V^P(Q^2,s,t) \, \varepsilon_{\mu \nu \alpha \beta} \,  k_1^{\nu} k_2^{\alpha}
k_3^{\beta} \, .
\end{equation}
Here the Mandelstam variables are defined as: $s=(Q-k_3)^2$, $t=(Q-k_1)^2$. In the following we will also use, for convenience, $u = Q^2 + 2 m_{\pi}^2 + m_P^2-s-t$.
\par
Neglecting the electron mass the cross-section of the $e^+ e^- \rightarrow \pi^+ \pi^- P$ process is given by:
\begin{equation} \label{eq:cx}
 \sigma_P(Q^2) = \frac{\alpha^2}{192 \, \pi \,Q^6} \, \int_{s_-}^{s_+} ds \int_{t_-}^{t_+} dt \, \phi(Q^2,s,t) \, |F_V^P(Q^2,s,t)|^2 ,
\end{equation}
where
\begin{equation} \label{eq:phiq2st}
 \phi(Q^2,s,t) = s t (Q^2-s-t)  +  s m_P^2(t-Q^2) -m_{\pi}^2 [m_P^4-m_P^2(2 Q^2+s)+Q^4-Q^2 s-2st] -sm_{\pi}^4\, ,
\end{equation}
being $m_P = m_{\pi},m_{\eta}$, depending on the final state\footnote{We use the average pion mass $m_\pi = (2m_{\pi^+}+ m_{\pi^0})/3$ in the matrix element calculation whereas the physical masses, both $m_{\pi^+}$ and $m_{\pi^0}$, are used to construct the physical phase space.}. In Eq.~(\ref{eq:cx}) the integration limits are:
\begin{eqnarray}\label{eq:intlimits}
 s_- &=& 4 m_{\pi}^2 \, , \nonumber \\
s_+ &=& (\sqrt{Q^2}-m_P)^2 \, , \nonumber \\
t_{\pm} &=& \frac{1}{4\, s} \left\{ \left( Q^2-m_P^2 \right)^2 - \left[ \lambda^{1/2}(Q^2,s,m_P^2) \mp \lambda^{1/2}(s,m_{\pi}^2,m_{\pi}^2)\right]^2 \right\} \, .
\end{eqnarray}
\subsection{Evaluation of the matrix amplitudes}
\begin{figure}[tb]
\begin{center}
\includegraphics[width=10.5cm]{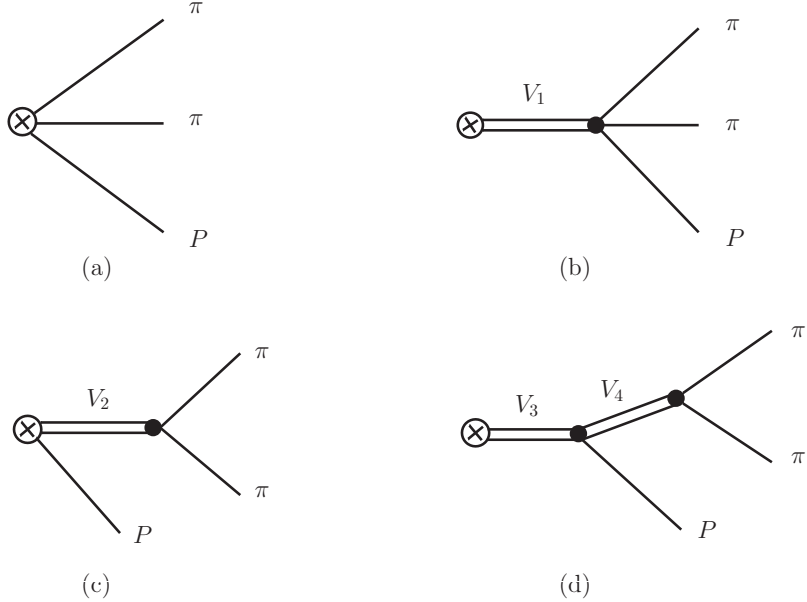}
\caption{\label{fig:1} Diagrams contributing to the hadronization of the vector current in $e^+ e^- \rightarrow \pi^+ \pi^- P$, $P=\pi^0, \eta$, at leading order in the $1/N_C$ expansion in the
isospin symmetric case. For $P = \pi^0$ we have $\boldsymbol{V} = [ (\omega, \phi), \rho, (\omega,\phi), \rho]$ for $i=1,2,3,4$, respectively. For $P = \eta$ we have $\boldsymbol{V} = [\rho, \rho, \rho, \rho]$. Here $\rho \equiv \rho(770)$, $\omega \equiv \omega(782)$ and $\phi \equiv \phi(1020)$. Diagram (a) is given by the Wess-Zumino anomaly, ${\cal L}^{\mbox{\tiny GB}}_{(4)}$ in 
Eq.~(\ref{eq:wzanomaly}).}
\end{center}
\end{figure}

The procedure to evaluate the amplitudes defined by Eq.~(\ref{eq:amplitude}) is now reduced to determining the vector form factor $F_V(Q^2,s,t)$ defined by Eq.~(\ref{eq:ffdef}),
through the hadronization of the vector current. Within the large-$N_C$ framework (with only one multiplet of vector resonances), one should consider all tree-level diagrams
generated by:
\begin{equation} \label{eq:rchtotal}
 {\cal L}_{\mbox{\tiny R$\chi$T}} = {\cal L}^{\mbox{\tiny GB}}_{(2)} + {\cal L}^{\mbox{\tiny GB}}_{(4)} + {\cal L}_{\mbox{\tiny V}} \,
\end{equation}
given by Eqs.~(\ref{eq:chpt2},\ref{eq:wzanomaly}) and (\ref{eq:lr}). The different topologies and particle content are specified in Figure~\ref{fig:1}.
For the sake of clarity we will discuss and show here the results for the vector form factors, in the $P=\pi^0,\eta$ cases, in a simplified treatment, where we assume
$\theta_P = 0$  in Eq.~(\ref{eq:etas}) and ideal vector mixing ($\sin \theta_V = 1/\sqrt{3}$) in Eq.~(\ref{eq:vectos}). We will also consider the vector resonances
to have zero width. Though this is not the proper physical framework it will be enough, on one side, to identify the main features of the final result, and on the other,
this simplified scheme is appropriate to settle information from QCD on the relevant couplings, as we will see in the next Section.
\par
At energies $E_{\mbox{\tiny CM}} \ll M_{\rho}$, the hadronization of the vector current is driven
by the anomaly, diagram (a) in Figure~\ref{fig:1}. At higher energies, $E_{\mbox{\tiny CM}} \sim 1 \, \mbox{GeV}$, vector resonances are needed to fulfil the dynamics of
the hadronization, as in diagrams (b), (c) and (d). The final result is given by the different diagram contributions as:
\begin{equation} \label{eq:fvabcd}
 F_V^P(Q^2,s,t) = F_{\mbox{\footnotesize a}}^P + F_{\mbox{\footnotesize b}}^P + F_{\mbox{\footnotesize c}}^P + F_{\mbox{\footnotesize  d}}^P, \; \; \; \; \; \; \; \; P = \pi, \eta \, ,
\end{equation}
where:
\begin{eqnarray} \label{eq:fvpion}
F_{\mbox{\footnotesize a}}^{\pi} &=&-\frac{N_C}{12 \pi^2 F^3}, \nonumber \\
F_{\mbox{\footnotesize b}}^{\pi}&=&\frac{ 8 \sqrt{2} F_V }{3 M_V F^3 (M_{\omega}^2-Q^2 )}
\{ (g_1-g_2-g_3) (Q^2-3m_{\pi}^2)+ 3 (2g_4+g_5) m_{\pi}^2\}   , \nonumber \\
F_{\mbox{\footnotesize c}}^{\pi}&=&-\frac{ 4 \sqrt{2} G_V }{3 M_V F^3 Q^2 (M_{\rho}^2-s )}
 \{c_1(Q^2+m_\pi^2-s)-c_2(Q^2-m_\pi^2-s)+8c_3 m_\pi^2  \nonumber\\
&&  \qquad \qquad \qquad \qquad  \qquad + \, c_5(Q^2-m_\pi^2+s)-2c_6 s \} + \{s\rightarrow t \}+\{s\rightarrow u \} , \nonumber\\
F_{\mbox{\footnotesize d}}^{\pi}&=&\frac{ 8  G_V F_V  }{3  F^3 (M_{\rho}^2-s ) (M_{\omega}^2-Q^2 )}
\{(d_1+8d_2)m_{\pi}^2+d_3(Q^2-m_\pi^2+s)\} \nonumber \\
&& \qquad \qquad \qquad \qquad \qquad \; \; \;  + \, \{s\rightarrow t \}+\{s\rightarrow u \},
\end{eqnarray}
and
\begin{eqnarray} \label{eq:fveta}
F_{\mbox{\footnotesize a}}^{\eta}&=&-\frac{N_C}{12 \sqrt{3}\pi^2 F^3} , \nonumber  \\
F_{\mbox{\footnotesize b}}^{\eta}&=&\frac{ 8 \sqrt{6}  F_V }{3 M_V F^3 (M_{\rho}^2-Q^2 )}
\{ (g_1-g_3)(s-2m_{\pi}^2)+g_2(2s-Q^2+m_\eta^2-2m_\pi^2) \nonumber \\
&& \qquad \qquad \qquad \qquad  \; \;  + \, (2g_4+g_5)m_\pi^2\},  \nonumber  \\
F_{\mbox{\footnotesize c}}^{\eta}&=&-\frac{ 4 \sqrt{6} G_V}{3 M_V F^3 (M_{\rho}^2-s )}
\{c_1(Q^2+m_\eta^2-s)-c_2(Q^2-m_\eta^2-s)+8c_3 m_\pi^2 \nonumber \\
&& \qquad \qquad \qquad \qquad \; \;  + \, c_5(Q^2-m_\eta^2+s)-2c_6 s \}, \nonumber \\
F_{\mbox{\footnotesize d}}^{\eta}&=&\frac{ 8 \sqrt{3} G_V F_V }{3  F^3 (M_{\rho}^2-s ) (M_{\rho}^2-Q^2 )}
\{d_1m_{\eta}^2+8d_2 m_{\pi}^2+d_3(Q^2-m_\eta^2+s)\}.
\end{eqnarray}
Our next step is to see how we can get information
about the coupling constants that appear in the vector form factors.

\section{Short-distance constraints on the resonance couplings}
\label{sec:4}

Our Lagrangian ${\cal L}_{\mbox{\tiny R$\chi$T}}$ in Eq.~(\ref{eq:rchtotal}) does not provide an effective theory of QCD for arbitrary values of its couplings.
The full theory should unambiguously predict the values of the coupling constants of the phenomenological Lagrangian. Unfortunately we still do not know how to
achieve this from first principles. However several ideas based on matching procedures have been developed and proven successful \cite{Ecker:1989yg,Knecht:2001xc,Pich:2002xy}.
One of these amounts to taking the rather reasonable assumption that the resonance region provides a bridge between the chiral and the perturbative regimes
\cite{Ecker:1989yg}. It is well known that the couplings of an effective theory valid in an energy range encode information from higher energy dynamics that
has been integrated out, hence the proposal involves matching the high-energy behaviour of Green functions (or related form factors) evaluated within the resonance
theory with the asymptotic results of perturbative QCD.
\par
A well known result from parton dynamics \cite{Brodsky:1973kr,Lepage:1980fj} indicates that vector and axial-vector form factors should behave smoothly at
high $Q^2$. This assessment is also consistent with the result that spectral functions associated to two-point Green functions of vector and axial-vector
currents go to a constant value at infinite momentum transfer \cite{Floratos:1978jb}. According to a local duality interpretation, the imaginary part of the quark loop, contributing to that spectral function, can be understood as the sum of infinite positive contributions of intermediate hadronic states. It is
therefore an educated guess that each of them should vanish in that limit if we want to fulfil the smoothness property.
\par
Our concern here is with the two-point Green function of the vector current $\Pi_V(Q^2)$, and in particular with the contribution of the exclusive channel of three pseudoscalars,
$\Pi_V^{P}(Q^2)$ defined by the three body phase space integration:
\begin{equation} \label{eq:piv3d}
 \int  d\Pi_3 T_{\mu}^{P} \, T_{\nu}^{P*}  = \left( Q^2 g_{\mu \nu} - Q_{\mu} Q_{\nu} \right) \, \Pi_V^{P}(Q^2) \, ,
\end{equation}
where $T_{\mu}^{P}$ has been defined in Eq.~(\ref{eq:amplitude}).
Hence:
\begin{equation} \label{eq:piv3}
 \Pi_V^{P}(Q^2) = \frac{\pi^2}{12 \, Q^4} \int  ds \, dt \, g^{\mu \nu} \, T_{\mu}^{P} \, T_{\nu}^{P*} \, ,
\end{equation}
and we impose the condition:
\begin{equation} \label{eq:const}
 \lim_{Q^2 \rightarrow + \infty} \Pi_V^{P}(Q^2) =  \lim_{Q^2 \rightarrow + \infty} \frac{\pi^2}{48 \, Q^4} \, \int_{s_-}^{s_+} ds \int_{t_-}^{t_+} dt \, \phi(Q^2,s,t) \,
|F_V^{P}(Q^2,s,t)|^2 = 0  \, ,
\end{equation}
with the function $\phi(Q^2,s,t)$ given in Eq.~(\ref{eq:phiq2st}) and the integration limits are those of (\ref{eq:intlimits}).
Notice that as the couplings cannot depend on the masses of the pseudoscalar mesons, we are able to perform the integration in the chiral limit ($m_{\pi}=m_{\eta}=0$).
By taking the case $P= \pi$ and $P= \eta$ we get, from (\ref{eq:fvpion}, \ref{eq:fveta}) and
for the couplings defined in Eqs.~(\ref{eq:lv2},\ref{eq:lv4},\ref{eq:lvv2}):
\begin{eqnarray}
 c_1-c_2+c_5 &=& 0 \, , \label{eq:rescons11} \\
 c_1+c_2+8c_3-c_5 &=& 0 \, , \label{eq:rescons12}\\
c_1-c_2-c_5+2c_6 & = & - \frac{N_C}{96 \sqrt{2} \pi^2} \frac{M_V}{G_V} \, , \label{eq:rescons14} \\
 d_3 &=& - \frac{N_C}{192 \, \pi^2} \frac{M_V^2}{F_V G_V} \, ,\label{eq:rescons13} \\
 g_2 &=& \frac{N_C}{192 \, \sqrt{2} \, \pi^2} \frac{M_V}{F_V} \, , \label{eq:rescons4} \\
g_1 - g_3 &=& - \frac{N_C}{96 \, \sqrt{2} \, \pi^2} \frac{M_V}{F_V} \, . \label{eq:rescons5}
\end{eqnarray}
These conditions coincide with those obtained in Ref.~\cite{Dumm:2009kj} coming from the analyses of $\tau \rightarrow K K \pi \nu_{\tau}$. However the constraints in
Eqs.~(\ref{eq:rescons12}, \ref{eq:rescons13}) do not agree exactly with those obtained in Ref.~\cite{RuizFemenia:2003hm} though the difference is not large.
In this regard we point out that the relations in this
later reference, obtained by matching of the leading OPE expansion of the $\langle VVP \rangle$ Green function, do not reproduce  the right asymptotic behaviour of
related form factors.
\par
To those conditions we add constraints that have been obtained in other settings, by demanding that the two-pion vector form factor vanishes at high $Q^2$ \cite{Ecker:1989yg} or
in the study of the $\langle VVP \rangle$ Green function \cite{RuizFemenia:2003hm}:
\begin{eqnarray} \label{eq:rescons2}
F_V G_V &=& F^2 \, , \nonumber \\
4 c_3 + c_1 &=& 0 \, , \nonumber \\
d_1 + 8 d_2 -d_3 &=& \frac{F^2}{8 \, F_V^2}  \, .
\end{eqnarray}
In view of our comment above on this later reference, it is reasonable to doubt  the validity of the latter two relations in Eq.~(\ref{eq:rescons2}). However as can be
seen in our result for the form factors, their role is very minor because the couplings $c_3$, $d_1$ and $d_2$ always contribute with factors of the pseudoscalar masses. Therefore we will use also these constraints.

\section{Further improvements on the form factors}
\label{sec:5}
Our form factors in Eqs.~(\ref{eq:fvpion}, \ref{eq:fveta}), as they stand, do not correspond to physical amplitudes (the resonance poles have no widths) and  only involve the energy region dominated by the lightest resonance multiplet. We intend to improve those form factors taking into account further corrections, adding resonance widths and
heavier resonance contributions in order to have a more physical and realistic form factor able to describe the energy domain below $E \sim 2 \, \mbox{GeV}$.
Accordingly we intend to  include all that information
as we explain in this section.

\subsection{Isospin symmetry breaking}
In order to take into account the violation of isospin symmetry provided by the $\rho$-$\omega$ mixing it is enough to include a new mixing angle $\delta$:
\begin{equation} \label{eq:rwmixing}
 \left( \begin{array}{c}
          \overline{\rho}^0 \\ \overline{\omega}
        \end{array}
 \right)
= \left( \begin{array}{cc}
          \cos \delta & \sin \delta \\ -\sin \delta & \cos \delta
         \end{array}
 \right) \, \left( \begin{array}{c}
                    \rho^0 \\ \omega
                   \end{array}
\right).
\end{equation}
The symmetric case corresponds to $\delta = 0$. Due to the $| \overline{s} s \rangle$ dominant quark content
of the $\phi(1020)$ resonance the relevance of isospin symmetry breaking in this later case is rather minor and we will not consider it.

\subsection{Higher order correction to the resonance coupling to the vector current}
The term in $F_V$ in Eq.~(\ref{eq:lv2}) provides the coupling of vector resonances to the vector current. If we do not consider ideal mixing for the
vector mesons, there is a contribution of the $\phi(1020)$ to the diagrams (b) and (d) in Figure~\ref{fig:1} and, from our analyses, it seems that the experimental data ask for a sizeable deviation of $F_V$ in the coupling of the $\phi(1020)$ resonance to the vector current.  Hence we add the contribution of a higher order chiral tensor coupled to the vector resonance in the first term in ${\cal L}^{\mbox{\tiny V}}_{(2)}$ (\ref{eq:lv2}) in the even-intrinsic-parity set. These next-to-leading operators have
been worked out in Ref.~\cite{Cirigliano:2006hb} and it happens that there is only one operator to be added at this order:
\begin{equation} \label{eq:fvho}
 {\cal L}_{(4)}^{\mbox{\tiny  V (even)}} = \lambda_6^V \, \langle V_{\mu \nu} \left\{f_{+}^{\mu \nu}, \chi_+ \right\} \rangle \, .
\end{equation}
For convenience we will redefine $\lambda_6^V = \alpha_V F_V / M_V^2$ in the following. As can be seen the modification to $F_V$ introduced by this operator is proportional
to pseudoscalar masses:
\begin{eqnarray} \label{eq:fvcor}
 F_V  &\longrightarrow&   F_V \left( 1 + \frac{8 \sqrt{2} \, \alpha_V m_{\pi}^2}{M_V^2} \right), \; \; \; \; \; \; \; \;\; \; \; \;\; \; \; \;\; \; \mbox{for} \; \rho^0, \omega , \nonumber \\
 F_V  &\longrightarrow&   F_V \left( 1 + \frac{8 \sqrt{2} \, \alpha_V (2 m_K^2- m_{\pi}^2)}{M_V^2} \right), \; \; \; \; \mbox{for} \; \phi ,
\end{eqnarray}
and, accordingly, its contribution will be tiny for the $\rho^0$ and $\omega$ case, while it could induce a non-negligible correction to the coupling of the $\phi(1020)$ resonance
to the vector current. Hence we will keep the shift in the latter case only, even for a non-ideal mixing of the neutral vector states.

\subsection{Inclusion of further resonance spectrum}
\label{subsec:5.3}
The vector form factors constructed in Section~\ref{sec:3} contain the dynamics of the lightest multiplet of vector resonances with masses 
below $E \sim 1 \, \mbox{GeV}$. If we wish to
extend the validity of our approach to higher energies, namely up to around $E \sim 2 \, \mbox{GeV}$, we have to include heavier resonances in our setting.
In particular the two multiplets that include the $\rho(1450)$ ($V'$) and $\rho(1700)$ ($V''$).
It is clear that to proceed by enlarging the phenomenological Lagrangian to these multiplets would be rather cumbersome and useless, because it would  triple the number of unknown coupling constants.
\par
Rather we propose performing the following substitutions in the form factors of $e^+ e^- \rightarrow \pi \pi \pi$:
\begin{equation} \label{eq:ms}
 \frac{1}{M_V^2 - x} \; \; \; \longrightarrow  \; \; \; \;  \frac{1}{M_V^2 - x} +  \frac{\beta_{\pi}'}{M_{V'}^2 - x} +  \frac{\beta_{\pi}''}{M_{V''}^2 - x} \, ,
\end{equation}
with $\beta_{\pi}'$ and $\beta_{\pi}''$ unknown real parameters that take into account the corresponding strength of the coupling of the $V'$ and $V''$ multiplets, respectively, with
regard to the lightest multiplet of vector resonances $V$. For the $\eta \pi \pi$ final state we introduce similar substitutions with $\beta_{\eta}'$ and
$\beta_{\eta}''$. This generalization respects the dynamics driven by the new multiplets, as they contribute with the same topologies of
those in Figure~\ref{fig:1}. However the procedure also implies that $\beta_P'$ and $\beta_P''$, $P=\pi, \eta$, include the information of isospin breaking in Eq.~(\ref{eq:rwmixing}) and
the deviation of $F_V$ in Eq.~(\ref{eq:fvcor}). We expect that experimental data will support our parameterization. However from a theoretical point of view,
this is clearly an oversimplification that is difficult to avoid at the moment given our present knowledge of QCD in this energy region.

\subsection{Width of resonances}
The vector form factors given in Eq.~(\ref{eq:fvabcd}) contain resonances with zero widths and, in consequence, are unphysical. In fact in the energy regime we are
considering, intermediate mesons do resonate and it is therefore compulsory to introduce their widths in our analyses. The natural procedure is to include a complex mass term ($M(q^2)$)
in the propagator of the resonances and to identify $\mbox{Im} \, M^2 (q^2) = - M_V \Gamma_V(q^2)$:
\begin{equation} \label{eq:inw}
 \frac{1}{M_V^2 - x} \; \; \; \; \longrightarrow \; \; \; \; \frac{1}{M_V^2 - x - i M_V \Gamma_V(x)} \; .
\end{equation}
In the case of wide resonances (to be compared with their mass) it is important to consider the $q^2$ dependence of the width. Our study requires us to consider the widths of the $\rho(770)$, $\omega(782)$,
$\phi(1020)$ and the corresponding heavier partners of the two additional multiplets. Our knowledge of the lightest vector resonances is very good and we know that while
the $\rho(770)$ is wide, both $\omega(782)$ and $\phi(1020)$ are very narrow therefore, for the later two, the off-shell behaviour is not going to be relevant.
Our knowledge of the two heavier vector multiplets is worse. We notice that both $\rho' \equiv \rho(1450)$ and $\rho'' \equiv \rho(1700)$ are rather wide and the $\omega'$ and the $\phi''$ are relatively narrow, while the $\omega ''$ is slightly wider. For definiteness we will consider the $\omega(782)$ and $\phi(1020)$ and their heavier partners as narrow resonances with a constant width.
\par
The issue of the construction of vector resonance off-shell widths was addressed in
Ref.~\cite{GomezDumm:2000fz} where the dominant contribution to the width of the $\rho(770)$, $\Gamma_{\rho}(q^2)$, was included:
\begin{equation} \label{eq:wrho}
 \Gamma_\rho(q^2)=\frac{M_\rho \,  q^2}{96\pi F^2}\left[\left(1-\frac{4m_\pi^2}{q^2}\right)^{3/2}\ \theta(q^2-4m_\pi^2)+\frac{1}{2}\left(1-\frac{4m_K^2}{q^2}\right)^{3/2}\ \theta(q^2-4m_K^2)\right] \, ,
\end{equation}
with $\theta(x)$ the step function.
Analogous results for the $\rho'$ and $\rho''$ have not been obtained, as we have poorer information on these resonances. In consequence
we have employed a reasonable parameterization, in terms of the on-shell widths, driven by the two-body phase space structure of the decay:
\begin{eqnarray} \label{eq:wprimas}
 \Gamma_{V}(q^2)&=&\Gamma_{V}(M_{V}^2) \frac{M_{V}}{\sqrt{q^2}}
\left[ \left(1-\frac{4m_\pi^2}{q^2}\right)/\left(1-\frac{4m_\pi^2}{M_{V}^2}\right)\right]^{3/2}\ \theta(q^2-4m_\pi^2),
\end{eqnarray}
for $V= \rho', \rho''$.

\subsection{Vector form factors improved}
In Appendix~\ref{app:A} we collect our final results for the form factors $F_V^{\pi}(Q^2,s,t)$ and $F_V^{\eta}(Q^2,s,t)$. Here we have taken into account
the corrections above on the initial results of Eqs.~(\ref{eq:fvpion},\ref{eq:fveta}). Moreover we give the expressions for general mixing angles $\theta_V$ and
$\theta_P$. Though rather cumbersome, they represent a sound parameterization of the
most relevant effects that are involved in the dynamics of these processes.
\par
These are the results that we will use in Section~\ref{sec:6} to analyse experimental data on the cross-sections.

\section{Decay widths involving vector resonances}
\label{sec:7}
The determination of decay widths implicating the lightest multiplet of vector resonances within our theoretical framework provides a parameterization
of those decays in terms of the couplings defined in Sections~\ref{sec:2} and \ref{sec:5} that, naturally, also appear in the cross-sections under consideration.
It is therefore relevant to take into account both, partial decay widths and cross-sections.
\par
We have calculated in our theoretical framework the physical decays $V \rightarrow \pi \pi$, $V \rightarrow \ell^+ \ell^-$, $V \rightarrow P \gamma$, $\eta' \rightarrow V \gamma$ and also
$V \rightarrow  \pi \pi \pi$, for $V = \rho(770),\,  \omega(782), \, \phi(1020)$ and $P= \pi,\, \eta$. The complete expressions are collected in Appendix~\ref{app:B}. Hence in Section~\ref{sec:6} we will analyze both cross-section and decay width data, the latter taken from
PDG 2012 \cite{Beringer:1900zz}. Notice that we do not include information coming from heavier resonance decays. On one side the experimental information of
decay widths coming from a second or third multiplet of vector resonances is rather loose. Moreover we only include the role of $\rho'$, $\omega'$, etc.,
through the shift in Eq.~(\ref{eq:ms}) and not through a phenomenological Lagrangian setting. Accordingly, we cannot evaluate those decays in the
present framework.
\par
As a consequence of this procedure our results for the Lagrangian couplings and mixing angles encode information from very wide and thorough sets of experimental data, while parameters involving heavier vector resonances could, in principle, be more uncertain.

\section{Fit of form factors and widths to experimental data}
\label{sec:6}
When analyzing data it is important to have an appropriate  -physics based (namely symmetries and dynamics of the underlying theories)-  
parameterization of the form factors. 
Our theoretical framework, as specified in the previous sections, provides a controlled setting that we can now use in order to extract information
from experimental data.
\par
We have employed a phenomenological Lagrangian based on our knowledge of the symmetries of hadrons and, in addition,
we have enforced a heuristic quantum field theory constraint on the vector-vector correlator as explained in Section~\ref{sec:4}. The output of our study
is a phenomenological symmetry-based parameterization of the form factors that contribute to the $e^+ e^- \rightarrow \pi \pi P$ cross-sections, with
$P= \pi,\eta$ and that we foresee would be a valid description for $E \lsim 2 \, \mbox{GeV}$. Unfortunately those form factors are not completely fixed from the theory side and our remaining ignorance is now parameterized in terms of
unknown coupling constants and mixing angles. We have to rely now on the analysis of the experimental data, namely the corresponding cross-sections and the decay widths involving vector resonances, to extract information on our unknowns.
\par
\begin{figure}[!]
\vspace*{-2.5cm}
\hspace*{-2cm}
\includegraphics[width=1.2\textwidth,height=1.15\textheight]{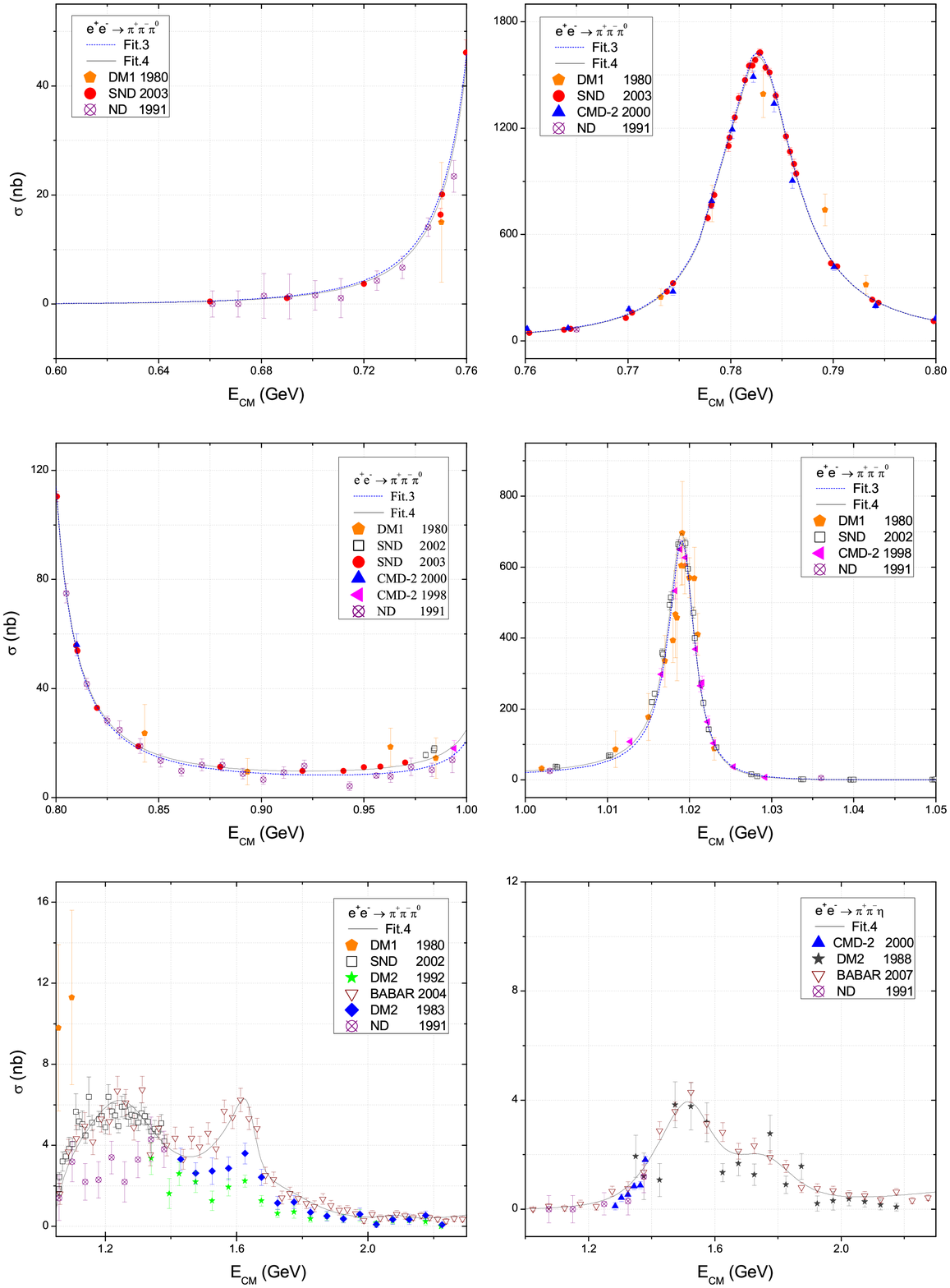}
\vspace*{-3.5cm}
\caption{\label{fig:2}  Comparison between the results of the fits (lines) explained in the text for the cross-sections for $e^+e^- \to \pi^+\pi^-\pi^0$  and  for $e^+e^- \to \eta \pi^+\pi^-$ (right-hand picture at bottom only). References for the experimental data are:  DM1 \cite{Cordier:1979qg}, DM2 \cite{Antonelli:1988fw}
ND \cite{Dolinsky:1991vq}, CMD-2 \cite{Akhmetshin:2003zn,Akhmetshin:1998se,Akhmetshin:2000ca}, SND \cite{Achasov:2003ir,Achasov:2002ud}, and BaBar \cite{Aubert:2004kj,Aubert:2007ef}.}
\end{figure}

Experimental data on $e^+ e^- \rightarrow \pi^+ \pi^- \pi^0$ have a long history \cite{Cordier:1979qg,Antonelli:1992jx,Achasov:2003ir,Akhmetshin:2003zn,Akhmetshin:1998se,Achasov:2002ud} that, until now, finishes with the BaBar data \cite{Aubert:2004kj} using radiative return. The $\eta \pi \pi$ final state has also been measured some time ago \cite{Dolinsky:1991vq,Antonelli:1988fw} and recently by BaBar \cite{Aubert:2007ef}. Experiments have focused on different energy regions as shown in Figure~\ref{fig:2}.
\par
We proceed to fit the experimental data with the use of MINUIT \cite{James:1975dr}. If not specified, all input parameters are taken from the
PDG 2012 \cite{Beringer:1900zz}. As a reference we fix the pion decay constant to $F = 0.0924$ GeV and we have used the constraints, given by
Eqs.~(\ref{eq:rescons11}-\ref{eq:rescons2}), in the form factors. The remaining unknown couplings are $F_V$, $2 g_4 + g_5$, $d_2$, $c_3$, $\alpha_V$ and the
phenomenological parameters $\beta_P$, $\beta_P'$ and $\beta_P''$ for $P=\pi,\eta$. The mixing angles $\theta_V$, $\theta_P$ and $\delta$ are also left free.
The parameters, listed appropriately in Table~\ref{tab:1}, are
fitted to the experimental data shown in Figure~\ref{fig:2}. We perform the following fits:
\begin{itemize}
\item[1/] {\bf Fit~1}: We consider both the $e^+ e^- \rightarrow \pi \pi \pi$ cross-section and the decay widths collected in the second column of Table~\ref{tab:2}.  We do not consider the isospin violating $\rho$-$\omega$ mixing ($\delta =0$) nor the $\alpha_V$ correction defined in relation with
Eq.~(\ref{eq:fvho}). Only the lightest multiplet of vector meson resonances is included in the analysis of the form factors and, therefore, the energy range of the cross-section analysed data runs from threshold up to $\sqrt{s} \sim 1.05 \, \mbox{GeV}$ only. Our results for the fitted parameters are those in the second columns in Tables~\ref{tab:1} and \ref{tab:2}.
\item[2/] {\bf Fit~2}: As in Fit~1 but we include the $\rho$-$\omega$ mixing in the analysis. The results are shown in the third columns in Tables~\ref{tab:1} and \ref{tab:2}.
\item[3/] {\bf Fit~3}: As in Fit~2 but we include the $\alpha_V$ parameter, according to Eq.~(\ref{eq:fvcor}), in the expressions for the form factors.
The results are shown in the fourth columns in Tables~\ref{tab:1} and \ref{tab:2}.
\item[4/] {\bf Fit~4}: We consider all decay widths and both the $e^+ e^- \rightarrow \pi \pi \pi$ and $e^+ e^- \rightarrow \eta \pi \pi$ cross-sections.
We choose to perform a global fit to both final states together because of the lack of experimental data
for the $\eta \pi \pi$ final state at threshold and in the very low-energy domain. The energy range of the analysed data reaches now $\sqrt{s} \sim 2.2 \, \mbox{GeV}$.
Accordingly we perform the shift explained in Subsection~\ref{subsec:5.3} and we include the three lightest multiplets of vector resonances whose masses and total widths are also fitted giving the results of the fifth columns in Tables~\ref{tab:1} and \ref{tab:2}.
\end{itemize}
The comparison of data on the cross-sections with our results for Fit~3 and Fit~4, where relevant, are shown in Figure~\ref{fig:2}. Notice that we do not
quote the corresponding values of the $\chi^2$ of our fits. Due to the huge variety of input data and different sources of errors we do not consider
the $\chi^2/\mbox{d.o.f.}$ as a representative parameter of how suitable the fits are.
\begin{table}[h!]
\begin{center}
\begin{tabular}{|c||c|c|c|c|c|}
\hline
                             & Fit 1      & Fit 2               & Fit 3             & Fit 4                  \\
\hline\hline
$F_V \, \mbox{(GeV)}$                 &  0.142$\pm$0.001      &   0.150$\pm$0.001       &  0.149$\pm$0.001   &     0.148$\pm$0.001    \\
$2g_4+g_5$                            & -0.395$\pm$0.002      &  -0.487$\pm$0.003       &  -0.489$\pm$0.003  &     -0.493$\pm$0.003     \\
$d_2$                                 &  0.0453$\pm$0.0022    &   0.0542$\pm$0.0019     &  0.0703$\pm$0.0023 &     0.0359$\pm$0.0007    \\
$c_3$                                 &  0.00837$\pm$0.00051  &   0.0114$\pm$0.0005     &  0.0156$\pm$0.0006 &     0.00689$\pm$0.00017  \\
$\alpha_V$                            &     -                 &   -                     &  0.0137$\pm$0.0007 &     0.0126$\pm$0.0007      \\
$\theta_V(^\circ)$                    &  39.45$\pm$0.01       &   39.26$\pm$0.01        &  38.84 $\pm$0.02   &     38.94$\pm$0.02     \\
$\theta_P(^\circ)$                    & -21.85 $\pm$0.41      &  -24.81 $\pm$0.35       &  -25.67$\pm$0.32   &     -21.37$\pm$0.26        \\
$\delta(^\circ)$                      &     -                 &   2.46$\pm$0.05         &    2.37$\pm$0.05   &     2.12$\pm$0.06        \\
$\beta_{\pi}'$                        &        -              &   -                     &  -                 &     -0.469$\pm$0.008      \\
$\beta_{\pi}''$                       &        -              &   -                     &  -                 &     0.225$\pm$0.007       \\
$\beta'_\eta$                         &        -              &   -                     &  -                 &     -0.174$\pm$0.017      \\
$\beta''_\eta$                        &        -              &   -                     &  -                 &     -0.0968$\pm$0.0139      \\
$M_{\rho'} \, \mbox{(GeV)} $          &        -              &   -                     &  -                 &     1.550$\pm$0.012      \\
$\Gamma_{\rho'} \, \mbox{(GeV)}$      &        -              &   -                     &  -                 &     0.238$\pm$0.018      \\
$M_{\omega'} \, \mbox{(GeV)}$         &        -              &   -                     &  -                 &     1.249$\pm$0.003      \\
$\Gamma_{\omega'} \, \mbox{(GeV)}$    &        -              &   -                     &  -                 &     0.307$\pm$0.007      \\
$M_{\phi '} \, \mbox{(GeV)}$          &        -              &   -                     &  -                 &     1.641$\pm$0.005      \\
$\Gamma_{\phi '}\, \mbox{(GeV)}$      &        -              &   -                     &  -                 &     0.086$\pm$0.007      \\
$M_{\rho''} \, \mbox{(GeV)} $         &        -              &   -                     &  -                 &     1.794$\pm$0.012      \\
$\Gamma_{\rho''} \, \mbox{(GeV)}$     &        -              &   -                     &  -                 &     0.297$\pm$0.033      \\
$M_{\omega''} \, \mbox{(GeV)}$        &        -              &   -                     &  -                 &     1.700$\pm$0.011      \\
$\Gamma_{\omega''} \, \mbox{(GeV)}$   &        -              &   -                     &  -                 &     0.400$\pm$0.013      \\
$M_{\phi ''}$ (GeV)                   &        -              &   -                     &  -                 &     2.086$\pm$0.022      \\
$\Gamma_{\phi ''}$ (GeV)              &        -              &   -                     &  -                 &     0.108$\pm$0.017    \\
\hline
\end{tabular}
\caption{\label{tab:1} Results for the different fits explained in the text.}
\end{center}
\end{table}
\begin{table}[h!]
\begin{center}
\begin{tabular}{|l||c|c|c|c|c|}
\hline
                                                     & Fit~1             & Fit~2        & Fit~3   &Fit~4   & PDG             \\
\hline\hline
$\Gamma_{\rho^0\to \pi\pi\pi} \; \; (10^{-5} \,  \mbox{GeV})$         &  -                &   1.23       &  1.15   & 0.93   &  1.51$\pm$1.19  \\
$\Gamma_{\omega\to \pi\pi\pi} \; \; (10^{-3} \, \mbox{GeV})$          &  6.67             &   7.46       &  7.53   & 7.66   &  7.57$\pm$0.06  \\
$\Gamma_{\phi\to  \pi\pi\pi} \; \; (10^{-4}\, \mbox{GeV})$            &  7.81             &   7.18       &  5.82   & 6.25   &  6.53$\pm$0.14  \\
$\Gamma_{\rho\to  ee} \; \; (10^{-6}\, \mbox{GeV})$                   &  5.80             &   6.69       &  6.68   & 6.54   &  7.04$\pm$0.06  \\
$\Gamma_{\omega\to  ee} \; \; (10^{-7}\, \mbox{GeV})$                 &  7.72             &   6.67       &  6.60   & 6.69   &  6.00$\pm$0.20  \\
$\Gamma_{\phi\to  ee} \; \; (10^{-6}\, \mbox{GeV})$                   &  0.88             &   0.99       &  1.24   & 1.20   &  1.25$\pm$0.01  \\
$\Gamma_{\rho\to  \pi\pi} \; \; (10^{-1}\, \mbox{GeV})$               &  1.24             &   1.10       &  1.12   & 1.14   &  1.49$\pm$0.01  \\
$\Gamma_{\omega\to  \pi\pi} \; \; (10^{-4}\, \mbox{GeV})$             &  -                &   2.12       &  1.99   & 1.61   &  1.30$\pm$0.10  \\
$\Gamma_{\phi\to  \pi\pi} \; \; (10^{-7}\, \mbox{GeV})$               &  1.95             &   2.19       &  2.76   & 2.66   &  3.15$\pm$0.55  \\
$\Gamma_{\rho^0\to  \pi^0 \gamma} \; \; (10^{-5} \, \mbox{GeV})$      &  4.45             &   6.35       &  6.21   & 5.96   &  8.95$\pm$1.19  \\
$\Gamma_{\rho^+\to  \pi^+ \gamma} \; \; (10^{-5} \, \mbox{GeV})$      &  4.42             &   4.97       &  4.90   & 4.81   &  6.71$\pm$0.75  \\
$\Gamma_{\omega\to  \pi^0 \gamma} \; \; (10^{-4} \, \mbox{GeV})$      &  4.14             &   4.51       &  4.50   & 4.43   &  7.03$\pm$0.23  \\
$\Gamma_{\phi\to  \pi^0\gamma} \; \; (10^{-6}\, \mbox{GeV})$          &  8.67             &   8.88       &  7.08   & 7.34   &  5.41$\pm$0.26  \\
$\Gamma_{\rho\to  \eta\gamma} \; \; (10^{-5}\, \mbox{GeV})$           &  4.66             &   5.77       &  5.96   & 4.85   &  4.47$\pm$0.30  \\
$\Gamma_{\omega\to  \eta\gamma} \; \; (10^{-6} \, \mbox{GeV})$        &  5.23             &   4.83       &  5.02   & 4.13   &  3.90$\pm$0.34  \\
$\Gamma_{\phi\to  \eta\gamma} \; \; (10^{-5} \, \mbox{GeV})$          &  5.87             &   5.98       &  6.04   & 6.57   &  5.58$\pm$0.10  \\
$\Gamma_{\eta'\to  \rho \gamma} \; \; (10^{-5} \, \mbox{GeV})$        &  5.51             &   5.53       &  5.50   & 5.37   &  5.68$\pm$0.10  \\
$\Gamma_{\eta'\to \omega\gamma} \; \; (10^{-6} \, \mbox{GeV})$        &  6.69             &   5.06       &  4.81   & 5.12   &  5.34$\pm$0.43  \\
$\Gamma_{\phi\to  \eta' \gamma} \; \; (10^{-7} \, \mbox{GeV})$        &  2.77             &   2.82       &  2.83   & 3.93   &  2.66$\pm$0.09  \\
\hline
\end{tabular}
\caption{\label{tab:2} Decay widths involving vector resonances. PDG data refer to \cite{Beringer:1900zz}.}
\end{center}
\end{table}
%
\begin{table}[h!]
\begin{center}
\begin{tabular}{|c||c|c|c|c|}
\hline
& $M_{\mbox{\tiny V',V''}}$ (GeV)    & $M_{\mbox{\tiny PDG}}$ (GeV)    & $\Gamma_{\mbox{\tiny V',V''}}$ (GeV)   & $\Gamma_{\mbox{\tiny PDG}}$ (GeV)    \\
\hline\hline
$\rho$(1450) &  1.550 & 1.465(25) & 0.238 & 0.400(60) \\
$\omega$(1420) & 1.249 &  1.40--1.45 & 0.307 & 0.18--0.25 \\
$\phi$(1680) & 1.641 &  1.680(20) & 0.086 & 0.15(5) \\
$\rho$(1700) & 1.794 & 1.720 (20) & 0.297 & 0.25(10) \\
$\omega$(1650)& 1.700 & 1.670(30) & 0.400 &0.315(35)\\
$\phi$(2170) & 2.086 & 2.175(15) & 0.108 & 0.061(18) \\
\hline
\end{tabular}
\caption{\label{tab:3} Comparison between the values of masses and total widths fitted for the $V'$ and $V''$ states ($P= \rho, \, \omega, \, \phi$)
and the seemingly corresponding ones in the PDG \cite{Beringer:1900zz}.}
\end{center}
\end{table}
\subsection{Analysis of results}
Tables~\ref{tab:1} and \ref{tab:2} collect all the results of our phenomenological analysis of the experimental data. The errors in
the fitted parameters in Table~\ref{tab:1} are provided by the MINUIT code and do not include the incertitude associated to our theoretical
framework. That is why we do not include errors in the partial widths of Table~\ref{tab:2}. We comment now on our results focusing
essentially in those provided by Fit~4:
\begin{itemize}
\item[-] We notice that the value that we get for the coupling $F_V$ is smaller by $\sim 20\%$ than the one obtained in the study of hadron tau
decays \cite{Dumm:2009va}. The inclusion of the $\alpha_V$ parameter in Fits~3 and 4 does not affect appreciably the fitted value of $F_V$.
\item[-] The fitted value for the combination of couplings $2 g_4 + g_5$ is very stable in all fits, 
around  $2 g_4 + g_5 \sim -0.5$. This is $20\%$ smaller than the value obtained from tau decays \cite{Dumm:2009kj}: $2 g_4 + g_5 = -0.60\pm0.02$.  
Notice, in addition, that  $2 g_4 + g_5$ is almost two orders of magnitude larger than $g_2 \simeq 0.0059$ as given by  Eq.~(\ref{eq:rescons4}). 
Meanwhile the values of $c_3$ and $d_2$ change largely from Fits~3 to 4.  Notice that the three couplings, $2 g_4+g_5$, $d_3$ and $c_2$ appear, for isospin conserving amplitudes, always suppressed by the squared of the pion mass, as can be seen in Eqs.~(\ref{eq:fvpion}, \ref{eq:fveta}). However when
$\eta-\eta'$ and $\phi-\omega-\rho^0$ mixings are considered, both $c_3$ and $d_2$ appear multiplying $m_K^2$ too, in the rather cumbersome expression of the form factor $F_V^{\eta}(Q^2,s,t)$  in Eq.~(\ref{eq:aeta}). This could be at the origin of the different behaviour of the Lagrangian coupling constants in the fit.
\item[-] The results for the mixing angles are the ones expected. On one side $\theta_V$ approaches the ideal mixing case $\theta_V = 35^{\circ}$.
The $\eta-\eta'$ mixing, defined in Eq.~(\ref{eq:etas}), $\theta_P = -21.4 \pm 0.3$ is very near to the one arising in the large-$N_C$ analyses \cite{HerreraSiklody:1997kd,Kaiser:1998ds}, and finally the isospin violating angle $\delta \sim 2^{\circ}$ is very small.
\item[-] With respect to the relative weight of the two heavier multiplets we notice that
$\beta_{\eta}'/ \beta_{\eta}'' \sim - \beta_{\pi}'/ \beta_{\pi}'' \simeq 2$. They contribute constructively in the $\eta$ case but with an
opposite relative sign in the three pions cross-section. With respect to the leading lightest multiplet of resonances we find $\beta_{\pi}' \simeq -0.5$
and $\beta_{\eta}' \simeq -0.2$.
\item[-] The results obtained for masses and total widths of the two heavier multiplets of vector resonances, namely $V'$ and $V''$, for
$V = \rho, \, \omega, \,  \phi$, resemble rather well the spectrum spread in PDG 2012 \cite{Beringer:1900zz}. We compare them in Table~\ref{tab:3}. It is interesting to notice that though we identify our $\rho'$ with
the well known $\rho$(1450), our results for the mass and total width most resemble those of the not-so-well-known $\rho$(1570) that has
$M = 1.57(7)$ (GeV) and $\Gamma = 0.144 (90)$ (GeV). It is important to point out, though, that the masses obtained in our fit correspond to
the definition given in Eq.~(\ref{eq:inw}), that is not necessarily that used by the experimentalists. From a quantum field theory
approach the pole in that equation should be at $M_V^2(x)-x-iM_V(x)\Gamma_V(x)$, but we take $M_V(x) \simeq M_V$. As a consequence a precise
comparison with the experimental determinations is not straightforward. Here we only comment on the general features of our results.
\item[-] A look to Table~\ref{tab:2} shows that our Fit~4 provides overall a reasonable approximation to the partial widths that involve
vector resonances. It is curious to notice that (maybe with the exception of the $\phi \rightarrow \eta' \gamma$ decay) Fit~3 does not give
better results than Fit~4. One could expect that focusing in the low-energy region of the cross-section where the lightest multiplet of
resonances lies as for Fit~3, it would give a more accurate prescription for the decay widths involving those vector states. However Fit~4
resembles very much Fit~3. 
The reason is that including heavier resonances as intermediate states would give a better description of the cross section, even in the low energy region.
The inclusion in Fit~4 of the  $e^+ e^- \rightarrow \eta \pi \pi$ cross-section data could also be the reason for the
better results for decay widths involving $\eta$ and $\eta'$ (with the exception of $\phi \rightarrow \eta \gamma$ and $\phi \rightarrow \eta' \gamma$).
\item[-] In Figure~\ref{fig:2} we compare our description from the fit with the experimental data. In the low-energy region
($E_{cm}  \lesssim 1.05 \, \mbox{GeV}$) we show both Fit~3 and Fit~4 results, that are almost indistinguishable.
However Fit~4 is a little better in the $0.8-1.05 \, \mbox{GeV}$ region. It convinces us that including heavier resonances as intermediate states 
improves definitely our approach. For the high-energy region
(bottom panels) we show only Fit~4 results. The left bottom panel presents $\sigma(e^+ e^- \rightarrow \pi \pi \pi)$ and the right bottom panel presents
$\sigma(e^+ e^- \rightarrow \eta \pi \pi)$.
\end{itemize}

\section{Implementation of our results in PHOKHARA 7.0}
\label{sec:8}
\begin{figure}[!]
\begin{minipage}[h]{0.6\textwidth}
\centering {\hspace{-2.5cm}
\includegraphics[width=0.8\textwidth,height=0.9\textwidth]{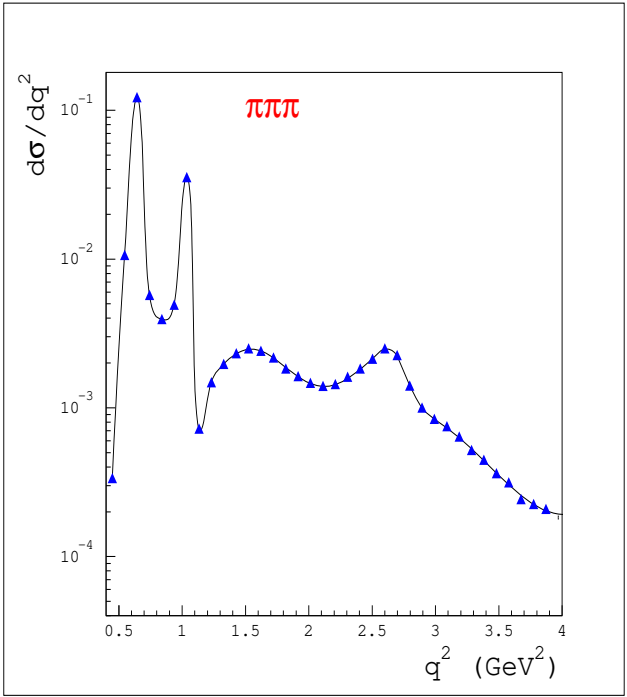}}
\end{minipage}
\begin{minipage}[h]{0.6\textwidth}
\centering {\hspace{-4.5cm}
\includegraphics[width=0.8\textwidth,height=0.9\textwidth]{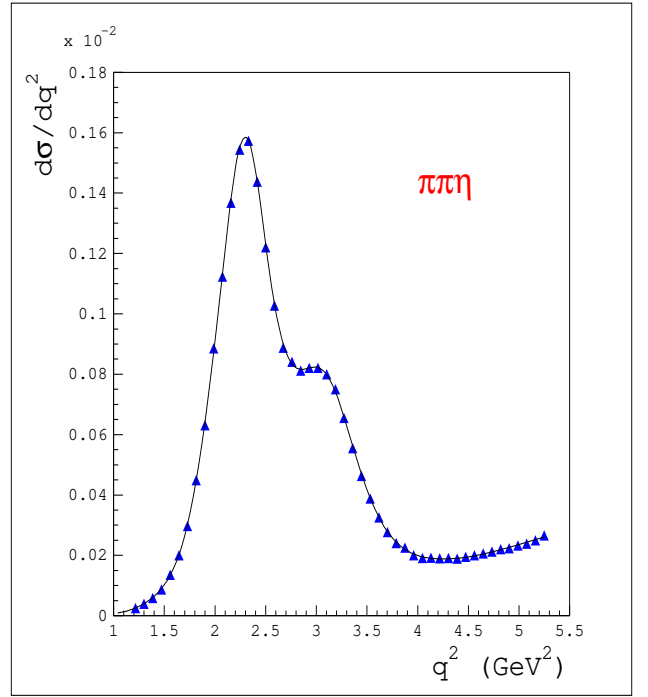}}
\end{minipage}
\vspace*{0cm}
\caption{\label{fig:3} Comparison between the Monte Carlo result (blue triangles) with R$\chi$T form factors and the analytical one (solid line) for both channels: $e^+e^-\to \pi^+\pi^-\pi^0$ (left) and $e^+e^-\to \pi^+\pi^-\eta$ (right). }
\end{figure}
\begin{figure}[!]
\begin{minipage}[h]{0.6\textwidth}
\centering {\hspace{-2.5cm}
\includegraphics[width=0.8\textwidth,height=0.9\textwidth]{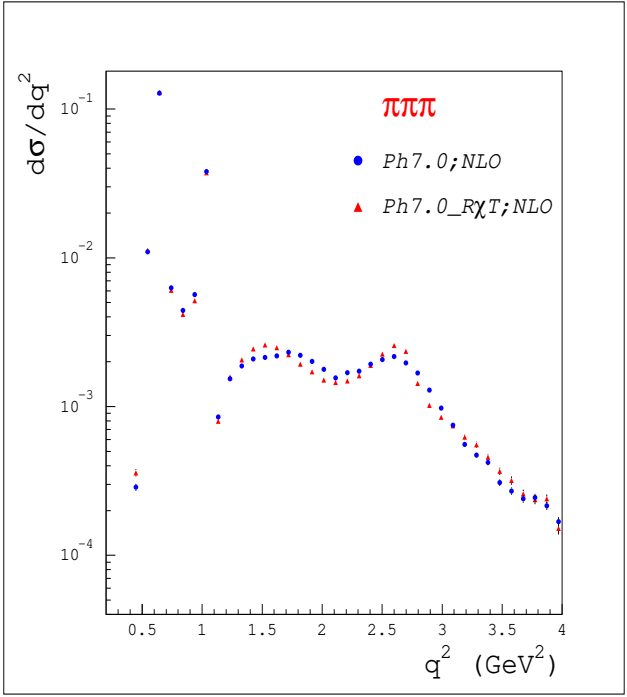}}
\end{minipage}
\begin{minipage}[h]{0.6\textwidth}
\centering {\hspace{-4.5cm}
\includegraphics[width=0.8\textwidth,height=0.9\textwidth]{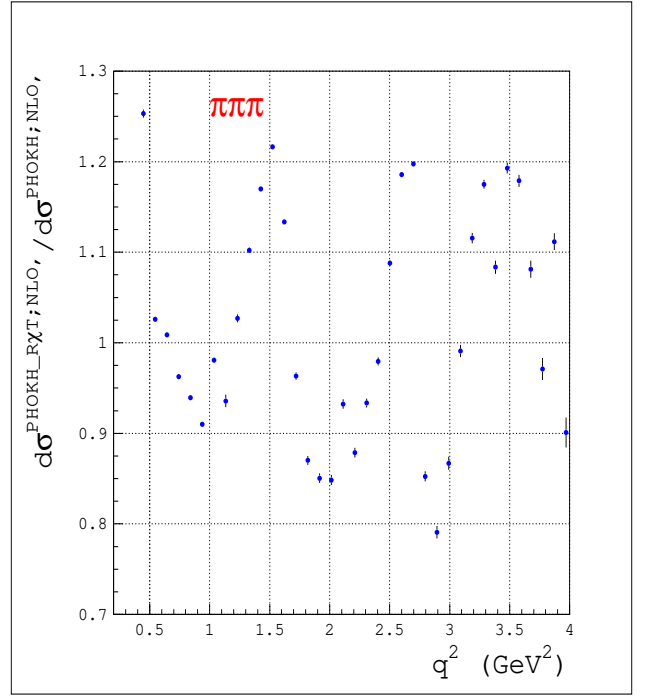}}
\end{minipage}
\vspace*{0cm}
\caption{\label{fig:4} The results from PHOKHARA7.0 for the three pion mode for both versions of the form factors: R$\chi$T and the original one, using the NLO setting.}
\end{figure}
PHOKHARA is a Monte Carlo event generator for the study of hadron and lepton production in electron positron annihilation at low energies
\cite{Rodrigo:2001kf,Czyz:2002np,Czyz:2003ue,Czyz:2004ua,Czyz:2004rj,phokhara_cite}. In Ref.~\cite{Czyz:2005as} the $\pi^+ \pi^- \pi^0$ final state was included in the code using vector meson dominance. Though this parameterization  provided a good agreement with
data, it did not include relevant QCD information that we try to realize with this article.
\par
We have implemented our form factors in both $\sigma(e^+ e^- \rightarrow \pi \pi \pi)$ and $\sigma(e^+ e^- \rightarrow \eta \pi \pi)$ in the PHOKHARA 7.0 version \cite{phokhara_cite}. The latter process has been added to this code for the first time and its pre-sampler distribution is the same as for the
three pion mode. In order to check the validity of our implementation, we perform a test of the form factors by comparing the leading-order (LO) 
 MC results (only one hard photon is emitted from the initial state),  at  the configuration
without any cuts on both pion and photon phase space, with the
analytical ones~\footnote{In fact, the same comparison has to be done also for the next-to-leading (NLO) version of the code. However, we restrict ourselves to the LO level as only the result for the hadronic form factor has been replaced in the code.}.
The ISR correction due to one hard photon emission changes the hadronic invariant mass ($q^2$) differential distribution as:
\begin{equation} \label{eq:1-pho}
\frac{d\sigma_{\mbox{\tiny ISR}}}{d q^2} = \frac{\alpha}{\pi} \,  \sigma_{\mbox{\tiny no-}\gamma}(q^2)\frac{Q^4+q^4}{Q^4(Q^2-q^2)} \,  \left(\ln{\frac{Q^2}{m_e^2}}-1\right) \, ,
\end{equation}
\begin{figure}[!]
\vspace*{-3cm}
\hspace*{-1cm}
\includegraphics[width=1.15\textwidth,height=1.1\textheight]{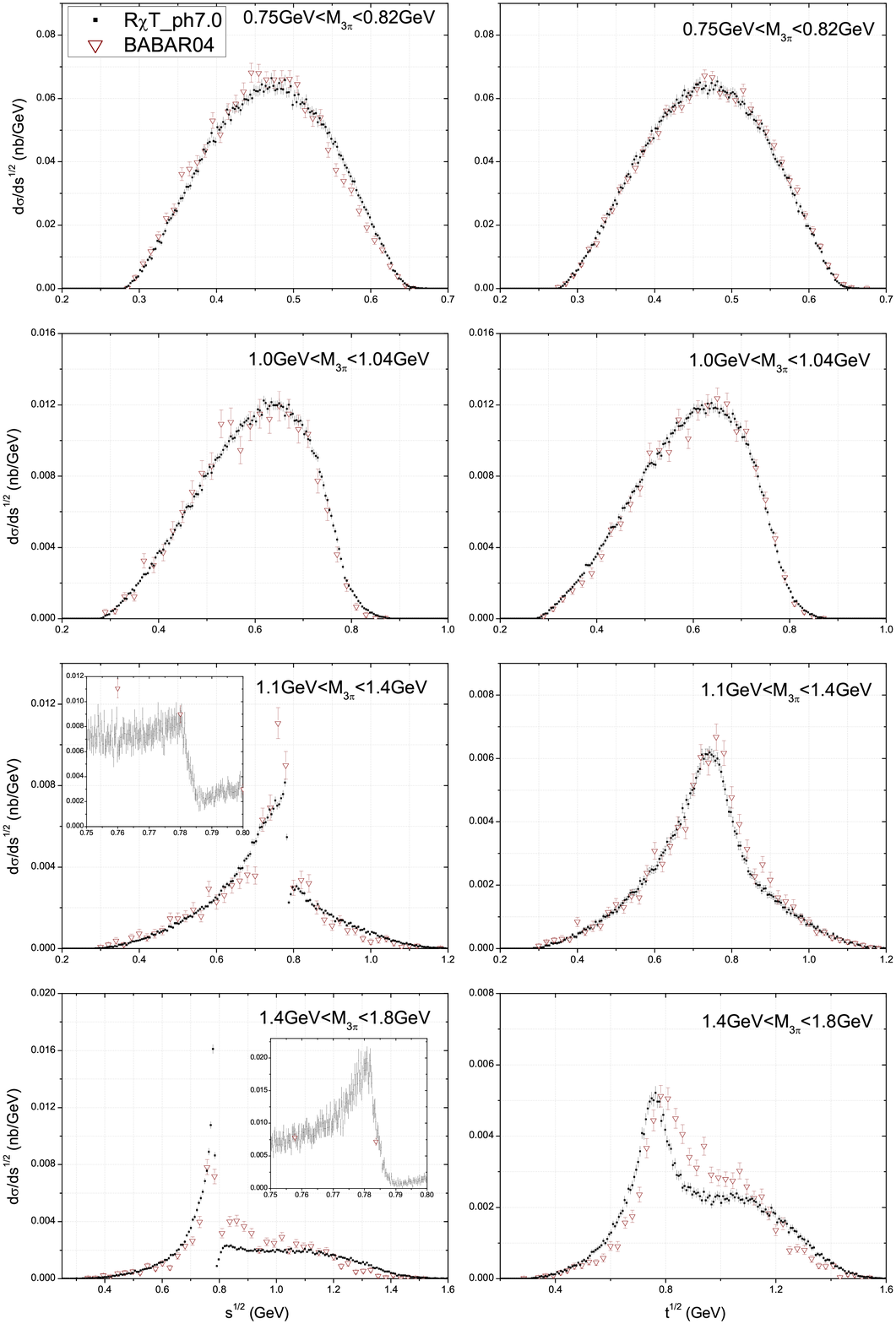}
\vspace*{-2cm}
\caption{\label{fig:5}  $\frac{d\sigma}{d M_{2\pi}}$ for $\sigma(e^+e^- \to\pi^+\pi^-\pi^0 \gamma)$ from PHOKHARA7.0 for $0.75 < M_{3 \pi} < 1.8$ GeV. Notice
that $M_{\pi^+\pi^-}$ ($\sqrt{s}$) and $M_{\pi^+\pi^0}$ ($\sqrt{t}$). BABAR04 data from \cite{Aubert:2004kj}. Detail of the $\rho$-$\omega$ interference effect
is also shown.}
\end{figure}
where $\sigma_{\mbox{\tiny no-}\gamma}(q^2)$ is our non-radiative result in Eq.~(\ref{eq:cx}).
The comparisons, for both $\pi \pi \pi$ and $\eta \pi \pi$ final states, are presented in Figure~\ref{fig:3}. As one can see the Monte Carlo prediction coincides very well with the analytical result.
\par
In  Figure~\ref{fig:4}, we present the MC results for three pion mode for both version of the form factor: R$\chi$T, and the original one \cite{Czyz:2005as}.
We run the NLO version. The difference between models is within 15\% in the region of validity of our approach, though we notice that this deviation is
most noticeable around $q^2 \sim 2.5 \, \mbox{GeV}^2$.
In Figure~\ref{fig:5}  the two pion invariant mass distributions, $M_{\pi^+\pi^-}$ ($\sqrt{s}$) and $M_{\pi^+\pi^0}$ ($\sqrt{t}$), for different
ranges of invariant masses of the three pions, are compared with data from BaBar~\cite{Aubert:2004kj}. Following Ref.~\cite{Czyz:2005as}, BaBar data points,
given as events/bin, are superimposed on plots obtained by PHOKHARA. We see that the agreement is rather good for $M_{3 \pi} \lesssim 1.4 \, \mbox{GeV}$,
while for larger values the concordance is slightly worse. In the distribution of two charged pions and for $1.1 \, \mbox{GeV} < M_{3 \pi} < 1.8 \, \mbox{GeV}$
it is possible to notice, around $\sqrt{s} \sim 0.8 \, \mbox{GeV}$, the effect of the $\rho-\omega$ interference as a dip in the spectrum.
\par
The PHOKHARA 7.0 code implemented with our form factors is available at the Downloads link of {\tt http://ific.uv.es/lhcpheno}.

\section{Conclusions}
\label{sec:9}
Low-energy hadron production in electron positron annihilation is the current setting at flavour factories like, for instance,  BaBar, KLOE, CMD,  Belle and
the future Belle II. At present the Monte Carlo code PHOKHARA 7.0 incorporates a good and relevant set of exclusive hadron
final states, including $\pi^+ \pi^- \pi^0$. However, in this 3$\pi$ case the parameterization that has been included can indeed 
be improved
as, for instance, it does not incorporate the anomalous contribution in that channel.
We believe that any correct implementation of the form factors that drives the hadronization of currents in the non-perturbative regime of QCD
should embody all the available information, like symmetries and dynamics, of QCD. Only in this case we can extract, by
phenomenological analysis of experimental data, valuable and reliable information on the hadronization properties.
\par
We have used the framework provided by Resonance Chiral Theory to work out the form factors relevant for the description of
$\sigma (e^+ e^- \rightarrow \pi^+ \pi^- \pi^0)$ and $\sigma (e^+ e^- \rightarrow \pi^+ \pi^- \eta)$ in the energy region from threshold until $E \lesssim 2 \, \mbox{GeV}$.
As the partial decay widths that involve vector resonances also depend on the couplings and parameters of R$\chi$T we have performed a joint analysis of the
cross-sections and several two-body, three-body and radiative decays that include vector resonances. Hence our results provide a wide view of all of them.
We find a generally good description of both cross-sections, from threshold to $E \sim 2 \, \mbox{GeV}$, and partial decay widths.
Our results for the cross-sections are shown in Figure~\ref{fig:2}.
\par
We have implemented our new description of the vector form factors for both $\pi^+ \pi^- \pi^0$ and $\pi^+ \pi^- \eta$ final states in the Monte Carlo
event generator PHOKHARA 7.0. The three pion case supersedes a
previous result, while for the $\pi \pi \eta$ final state this is the first implementation. The agreement of di-pion mass spectra that results
from the code is reasonably good.
\par
Hadronization is a key feature of non-perturbative QCD, which at present cannot be treated from first principles. However
we know how to implement QCD properties within phenomenological Lagrangians, as in Resonance Chiral Theory. This is a tool that has proven to be
effective, for instance, in the analyses of hadronization of QCD currents in tau lepton decays. We have shown that it provides a reasonable
setting to study three pseudoscalar meson production in electron positron annihilation. Maybe one day the lattice or some future QCD-based framework
will give us a solution for QCD in the low-energy regime. Until then we have to use QCD-based tools that prove to be useful phenomenologically in the implementation of hadronization procedures.

\section*{Acknowledgements}
Conversations with Germ\'an Rodrigo on the topic of this paper
are warmly acknowledged. We would like to thank Henryk Czy\.z for fruitful discussions on context of MC PHOKHARA 7.0.
We also wish to thank Michael R. Pennington for a careful reading of our manuscript and for his suggestions.
Lingyun Dai thanks CSC (China Scholarship Council) for their support.
This research has been supported in part by the funds of Polish National Science Centre
under decisions DEC-2012/04/M/ST2/00240 and DEC-2011/03/B/ST2/00107 (O.S.)
and by the Spanish Government and ERDF funds from the EU Commission [grants FPA2007-60323,
FPA2011-23778, CSD2007-00042 (Consolider Project CPAN)].
This paper has in part been authored by Jefferson Science Associates, LLC under U.S. DOE Contract No. DE-AC05-06OR23177.

\appendix
\renewcommand{\theequation}{\Alph{section}.\arabic{equation}}
\renewcommand{\thetable}{\Alph{section}.\arabic{table}}
\setcounter{equation}{0}
\setcounter{table}{0}

\section{Complete results for the $F_V^P(Q^2,s,t)$ form factors in $e^+ e^- \rightarrow \pi^+ \pi^- P$}
\label{app:A}
\subsection{$P = \pi^0$}
\label{sub:1}
The vector form factor relevant for the $e^+ e^- \rightarrow \pi^+ \pi^- \pi^0$ cross-section, defined by Eq.~(\ref{eq:ffdef}), is given by:
\begin{equation} \label{eq:apion}
 F_V^{\pi}(Q^2,s,t) = F_a^{\pi} + F_b^{\pi} + F_c^{\pi} + F_d^{\pi} \, ,
\end{equation}
where (see Subsection~\ref{sub:3} for notation):
\begin{eqnarray}\label{eq:piFormfactor}
{F}_{a}^{\pi}&=&-\frac{N_C}{12\pi ^2F^3}, \nonumber\\[10mm]
{F}_{b}^{\pi} &=&\frac{8\sqrt{2}F_V(1+8 \sqrt{2}\alpha _V\frac{m_\pi^2}{M_V^2})}{3 M_V F^3} (\sqrt{2}\cos\theta _V+\sin\theta_V) \,  G_{R_\pi}(Q^2)
\times \nonumber \\
&& \left\{ (\sin\theta_V\cos \delta -\sqrt{3}\sin \delta )\cos \delta \;  BW_{R}[\pi,\omega,Q^2] \right. \nonumber\\
&& \left. \;  + (\sin\theta_V\sin \delta +\sqrt{3}\cos \delta) \sin \delta \;  BW_{R}[\pi,\rho,Q^2] \right\} \nonumber \\
&&+ \frac{8\sqrt{2}F_V (1+8 \sqrt{2}\alpha _V\frac{2m_K^2-m_{\pi}^2}{M_V^2})}{3M_V F^3}\cos\theta_V
(\cos\theta _V-\sqrt{2}\sin\theta _V) \, BW_{R}[\pi, \phi,Q^2]~G_{R_\pi}(Q^2), \nonumber\\[10mm]
{F}_{c}^{\pi}&=&-\frac{4\sqrt{2}G_V}{3M_V F^3} \left\{ (\cos \delta +\sqrt{6}\cos\theta _V \sin \delta +\sqrt{3}\sin \delta \sin\theta _V)\cos \delta~
BW_{R}[\pi,\rho,s]~C_{R\pi}(Q^2,s) \right. \nonumber\\
&&\; \; \; \; \; \; \; \; \; \; \; \; \; \; \, + BW_{R}[\pi,\rho,t]~C_{R\pi}(Q^2,t)+ BW_{R}[\pi,\rho,u]~C_{R\pi}(Q^2,u)\nonumber\\
&& \; \; \; \; \; \; \; \; \; \; \; \; \; \; \, \left. - \left[ \sqrt{3} \cos \delta \left(\sqrt{2}\cos \theta _V+\sin \theta_V
\right)-\sin \delta \right]\sin \delta ~BW_{R}[\pi,\omega,s]~C_{R \pi}(Q^2,s) \right\} ,\nonumber\\[10mm]
{F}_{d}^{\pi}&=&\frac{8G_V F_V(1+8 \sqrt{2}\alpha _V\frac{m_{\pi }{}^2}{M_V^2}) }{ 3 F^3}  (\sqrt{2}\cos\theta_V+\sin\theta _V) \times \nonumber \\
&& \left\{ (\sin\theta_V\cos \delta -\sqrt{3}\sin \delta )
\cos \delta \cos 2 \delta \; BW_{RR}[\pi,\omega,\rho,Q^2,s]~D_{R\pi}(Q^2,s) \right. \nonumber\\
&&\; \; + (\sin\theta _V\cos \delta -\sqrt{3}\sin \delta )\cos \delta \; BW_{RR}[\pi,\omega,\rho,Q^2,t]~D_{R\pi}(Q^2,t)\nonumber\\
&&\; \; +(\sin\theta _V\cos \delta -\sqrt{3}\sin \delta )\cos \delta \;  BW_{RR}[\pi,\omega,\rho,Q^2,u]~D_{R\pi}(Q^2,u)\nonumber\\
&&\; \; + (\sin\theta_V\cos \delta -\sqrt{3}\sin \delta )\sin 2\delta \sin \delta \;  BW_{RR}[\pi,\omega,\omega,Q^2,s]~D_{R\pi}(Q^2,s)\nonumber\\
&&\; \; + (\sin\theta _V\sin \delta +\sqrt{3}\cos \delta )\sin 2\delta \cos \delta \;  BW_{RR}[\pi,\rho,\rho,Q^2,s]~D_{R\pi}(Q^2,s)\nonumber\\
&&\; \; + (\sin\theta _V\sin \delta +\sqrt{3}\cos \delta )\sin \delta  \; BW_{RR}[\pi,\rho,\rho,Q^2,t]~D_{R\pi}(Q^2,t)\nonumber\\
&& \; \; + (\sin\theta _V\sin \delta +\sqrt{3}\cos \delta )\sin \delta \; BW_{RR}[\pi,\rho,\rho,Q^2,u]~D_{R\pi}(Q^2,u)  \nonumber\\
&&\left. \; \; - (\sin\theta_V\sin \delta +\sqrt{3}\cos \delta ) \cos 2 \delta \sin \delta \; BW_{RR}[\pi,\rho,\omega,Q^2,s]~D_{R\pi}(Q^2,s) \, \right\}
\nonumber \\
&&+\frac{8G_V F_V(1+8 \sqrt{2} \alpha _V\frac{2m_K{}^2-m_{\pi }{}^2}{M_V^2})}{3 F^3}   (\cos\theta _V-\sqrt{2}\sin\theta _V) \, \cos\theta_V \times \nonumber \\
&& \left\{ \cos^2 \delta \; BW_{RR}[\pi,\phi,\rho,Q^2,s]~D_{R\pi}(Q^2,s)  +  \sin^2 \delta  \; BW_{RR}[\pi,\phi,\omega,Q^2,s]~D_{R\pi}(Q^2,s)\right. \nonumber\\
&&\left. \; \; +  \; BW_{RR}[\pi,\phi,\rho,Q^2,t]~D_{R\pi}(Q^2,t) + \; BW_{RR}[\pi,\phi,\rho,Q^2,u]~D_{R\pi}(Q^2,u) \right\} .
\end{eqnarray}

\subsection{$P = \eta$}
\label{sub:2}
The vector form factor relevant for the $e^+ e^- \rightarrow \pi^+ \pi^- \eta$ cross-section, defined by Eq.~(\ref{eq:ffdef}), is given by:
\begin{equation} \label{eq:aeta}
 F_V^{\eta}(Q^2,s,t) = F_a^{\eta} + F_b^{\eta} + F_c^{\eta} + F_d^{\eta} \, ,
\end{equation}
where (see Subsection~\ref{sub:3} for notation):

\begin{eqnarray}\label{etaFormfactor}
{F}_{a}^{\eta}&=&-\frac{N_C}{12\sqrt{3}\pi^2 F^3}(-\sqrt{2}\sin\theta_P +\cos\theta_P),\nonumber\\[10mm]
{F}_{b}^{\eta}&=&\frac{8\sqrt{6}F_V (1+8 \sqrt{2} \alpha _V\frac{m_{\pi}^2}{M_V^2})}{3M_V F^3}\left(\cos \delta +\frac{1}{\sqrt{3}}\sin \delta \sin\theta _V\right)
\cos \delta (-\sqrt{2}\sin\theta_P+\cos \theta_P)\nonumber\\
&&\;\;\;~ \times BW_{R}[\eta,\rho,Q^2]~G_{R\eta}(Q^2,s)\nonumber\\
&&- \frac{8\sqrt{6}F_V(1+8 \sqrt{2} \alpha _V\frac{m_{\pi}^2}{M_V^2})}{3M_V F^3}\left(-\sin \delta +\frac{1}{\sqrt{3}}\cos \delta \sin\theta _V\right) \sin \delta (-\sqrt{2}\sin\theta_P+\cos\theta_P)\nonumber\\
&&\;\;\;~ \times BW_{R}[\eta,\omega,Q^2]~G_{R\eta}(Q^2,s),\nonumber\\[10mm]
{F}_{c}^{\eta}&=&-\frac{4\sqrt{2} G_V}{3 M_V F^3} \cos \delta  \{\sqrt{3} \cos \delta  (\cos \theta_P-\sqrt{2} \sin \theta_P )+
\sin \delta  [\;\sqrt{2} \cos \theta_V \cos \theta_P\nonumber\\
&&\;\;\; \;\;\;\;\;\;\;\;\;\;\;\;~-\sin\theta_V(\cos \theta_P+\sqrt{2} \sin \theta_P )\;]\;\}BW_{R}[\eta,\rho,s]~C_{R \eta 1}(Q^2,s,m_{\eta}^2)\nonumber\\
&&- \frac{4\sqrt{2}G_V}{9M_V F^3}\cos \delta   \{4 \sin \delta  [\sqrt{2} \cos(\theta _V+\theta_P )
-2 \cos \theta_P \sin\theta _V+\cos\theta _V \sin \theta_P] m_K^2+\nonumber\\
&&\;\;\;\;\;\; \;\;\;\;\;\;\;\;\;\;\;\;~[3 \sqrt{3} \cos \delta  (\cos \theta_P-\sqrt{2} \sin \theta_P)-\sin \delta  (~\sqrt{2} \cos(\theta _V+\theta_P )
-5 \cos \theta_P \sin\theta _V\nonumber\\
&&\;\;\;\;\;\; \;\;\;\;\;\;\;\;\;\;\;\;~+4 \cos\theta _V \sin \theta_P~)] m_{\pi}^2\;\}~BW_{R}[\eta,\rho,s]~C_{R\eta 2}\nonumber\\
&&+\frac{4\sqrt{2}G_V}{3M_V F^3} \sin \delta \{\sqrt{3} \sin \delta  (-\cos \theta_P+\sqrt{2} \sin \theta_P)
+\cos \delta  [\sqrt{2} \cos\theta _V \cos \theta_P\nonumber\\
&&\;\;\;\;\;\; \;\;\;\;\;\;\;\;\;\;\;\;~-\sin\theta _V (\cos \theta_P+\sqrt{2} \sin \theta_P)\;]\;\}~BW_{R}[\eta,\omega,s]~C_{R\eta 1}(Q^2,s,m_{\eta}^2)\nonumber\\
&&+\frac{4\sqrt{2}G_V}{9M_V F^3} \sin \delta  \{4 \cos \delta  [\sqrt{2} \cos(\theta _V+\theta_P )
-2 \cos \theta_P \sin\theta _V+\cos\theta _V\sin \theta_P] m_K^2-\nonumber\\
&&\;\;\;~\;\;\; \;\;\;\;\;\;\;\;\;\;\;\;[3 \sqrt{3} \sin \delta  (\cos \theta_P-\sqrt{2} \sin \theta_P)+\cos \delta  (\sqrt{2} \cos(\theta _V+\theta_P )
-5 \cos \theta_P \sin\theta _V\nonumber\\
&&\;\;\;\;\;\; \;\;\;\;\;\;\;\;\;\;\;\;~+4 \cos\theta _V \sin \theta_P) ] m_{\pi }^2\}~BW_{R}[\eta,\omega,s]~C_{R\eta 2},\nonumber\\[10mm]
{F}_{d}^{\eta}&=&\frac{8F_V(1+8 \sqrt{2} \alpha _V\frac{m_\pi^2}{M_V^2}) G_V}{\sqrt{6} F^3 }\cos \delta
\left(\cos \delta +\frac{1}{\sqrt{3}}\sin \delta \sin\theta _V \right)\nonumber\\
&&\;\;\;~\times \{\cos^2\delta (\sqrt{2} \cos \theta_P-2 \sin \theta_P )+\sin^2\delta
[\cos \theta_P \sin \theta _V (4 \cos \theta _V-\sqrt{2} \sin \theta _V)-2 \sin \theta_P ]\}\nonumber\\
&&\;\;\;~\times BW_{RR}[\eta,\rho,\rho,Q^2,s]~D_{R\eta 1}(Q^2,s,m_{\eta}^2)\nonumber\\
&&+ \frac{2F_V(1+ 8 \sqrt{2} \alpha_V\frac{m_\pi^2}{M_V^2}) G_V}{3\sqrt{6} F^3 }\cos \delta \left(\cos \delta +\frac{1}{\sqrt{3}}\sin \delta \sin \theta_V\right)\nonumber\\
&&\;\;\;~ \times \bigg\{8 \sin^2\delta [\cos \theta_P (-3 \sqrt{2}+\sqrt{2} \cos2\theta _V+4\sin2\theta _V)+\nonumber\\
&&\;\;\;\; \; \; \; ~(-3+\cos2\theta _V+2 \sqrt{2} \sin2\theta _V) \sin \theta_P ]
m_K^2+[12 \cos^2\delta (\sqrt{2} \cos \theta_P-2 \sin \theta_P )+\nonumber\\
&&\;\;\;~\sin^2\delta (-9 \sqrt{2} \cos(2\theta _V-\theta_P )+18 \sqrt{2} \cos \theta_P
+7\sqrt{2} \cos(2\theta _V+\theta_P )-8 \sin(2 \theta _V+\theta_P )) ]  m_{\pi }^2 \bigg\}\nonumber\\
&&\;\;\;~\times BW_{RR}[\eta,\rho,\rho,Q^2,s]~~D_{R\eta 2}\nonumber\\
&&- \frac{8F_V(1+8 \sqrt{2} \alpha _V\frac{m_\pi^2}{M_V^2})G_V}{\sqrt{6} F^3 }\sin \delta \left(-\sin \delta +\frac{1}{\sqrt{3}}\cos \delta \sin \theta_V\right)\nonumber\\
&&\;\;\;~\times \bigg\{\cos \theta_P [\sqrt{2} \sin^2\delta+\cos^2\delta \sin \theta _V
(4 \cos \theta _V-\sqrt{2} \sin \theta _V)]-2 \sin \theta_P \bigg\}\nonumber\\
&&\;\;\;~\times BW_{RR}[\eta,\omega,\omega,Q^2,s]~D_{R\eta 1}(Q^2,s,m_{\eta}^2)\nonumber\\
&&- \frac{2F_V(1+8 \sqrt{2} \alpha_V\frac{m_\pi^2}{M_V^2}) G_V}{3\sqrt{6} F^3 }\sin \delta \left(-\sin \delta +\frac{1}{\sqrt{3}}\cos \delta \sin \theta_V\right)\nonumber\\
&&\;\;\;~\times \bigg\{8 \cos^2\delta [\cos \theta_P ( -3 \sqrt{2}+\sqrt{2} \cos2\theta _V+4\sin2\theta _V )+\nonumber\\
&&\;\;\;~ (-3+\cos 2 \theta _V+2 \sqrt{2} \sin 2 \theta _V) \sin \theta_P ]m_K^2+[12 \sin^2\delta (\sqrt{2} \cos \theta_P-2 \sin \theta_P )+\nonumber\\
&&\;\;\;~\cos^2\delta (-9 \sqrt{2} \cos(2 \theta _V-\theta_P )+18 \sqrt{2} \cos \theta_P
+7\sqrt{2} \cos(2 \theta _V+\theta_P )-8 \sin(2 \theta _V+\theta_P ))\big] m_{\pi }^2\bigg\}\nonumber\\
&&\;\;\;~\times BW_{RR}[\eta,\omega,\omega,Q^2,s]~D_{R\eta 2}\nonumber\\
&&+ \frac{2F_V(1+8 \sqrt{2} \alpha _V\frac{m_\pi^2}{M_V^2})G_V}{\sqrt{6} F^3 }\cos \delta \left(-\sin \delta +\frac{1}{\sqrt{3}}\cos \delta \sin \theta _V\right)\cos \theta_P \sin 2 \delta \nonumber\\
&&\;\;\;~\times \left(-3 \sqrt{2}+\sqrt{2} \cos 2 \theta_V+4 \sin2\theta _V\right)~BW_{RR}[\eta,\omega,\rho,Q^2,s]~D_{R\eta 1}(Q^2,s,m_{\eta}^2)\nonumber\\
&&+ \frac{2F_V(1+8 \sqrt{2} \alpha_V\frac{m_\pi^2}{M_V^2}) G_V}{3\sqrt{6} F^3 }\cos \delta \left(-\sin \delta +\frac{1}{\sqrt{3}}\cos \delta \sin \theta _V\right) \sin 2\delta \nonumber\\
&&\;\;\;~\times \bigg\{4( -3+\cos 2 \theta_V+2 \sqrt{2} \sin 2 \theta_V ) \sin \theta_P  (m_K^2-m_{\pi }^2)
\nonumber\\
&&\;\;\;~+\cos \theta_P ( -3 \sqrt{2}+\sqrt{2} \cos2\theta_V+4 \sin2\theta_V )(4 m_K^2-m_{\pi }^2)\bigg\}~BW_{RR}[\eta,\omega,\rho,Q^2,s]~D_{R\eta 2} \nonumber\\
&&- \frac{2F_V(1+8 \sqrt{2} \alpha_V\frac{m_\pi^2}{M_V^2}) G_V}{\sqrt{6} F^3 }\sin \delta \left(\cos \delta +\frac{1}{\sqrt{3}}\sin \delta \sin \theta _V\right) \cos \theta_P \sin2\delta \nonumber\\
&&\;\;\;~\left( -3 \sqrt{2}+\sqrt{2} \cos2\theta _V+4 \sin2\theta_V \right)~BW_{RR}[\eta,\rho,\omega,Q^2,s]~D_{R\eta 1}(Q^2,s,m_{\eta}^2)\nonumber\\
&&-\frac{2F_V(1+8 \sqrt{2} \alpha_V\frac{m_\pi^2}{M_V^2}) G_V}{3\sqrt{6} F^3 }\sin \delta \left(\cos \delta +\frac{1}{\sqrt{3}}\sin \delta \sin \theta _V\right) \sin2\delta\nonumber\\
&&\;\;\;~\bigg\{4 [ -3+\cos2\theta_V+2 \sqrt{2} \sin2\theta_V ] \sin \theta_P  (m_K^2-m_{\pi }^2)
\nonumber\\
&&\;\;\;~+\cos \theta_P [-3 \sqrt{2}+\sqrt{2} \cos2\theta_V+4 \sin2\theta_V ~]~(4 m_K^2-m_{\pi }^2)\bigg\} ~BW_{RR}[\eta,\rho,\omega,Q^2,s]~D_{R\eta 2}\nonumber\\
&&- \frac{4F_V(1+8 \sqrt{2} \alpha _V\frac{2m_K{}^2-m_\pi^2}{M_V^2}) G_V}{3\sqrt{2} F^3 }\cos \delta \cos \theta _V
\nonumber\\
&&\;\;\;~\times \cos \theta_P \sin \delta  \left( -4 \cos2\theta_V+\sqrt{2} \sin2\theta_V \right)~BW_{RR}[\eta,\phi,\rho,Q^2,s]~D_{R\eta 1}(Q^2,s,m_{\eta}^2)\nonumber\\
&&+ \frac{4F_V(1+8 \sqrt{2} \alpha_V\frac{2m_K^2-m_\pi^2}{M_V^2}) G_V}{9\sqrt{2} F^3 }\cos \delta \cos \theta _V \sin \delta \nonumber\\
&&\;\;\;~\times \bigg\{ 4 ( 2 \sqrt{2} \cos2\theta_V-\sin2\theta_V ) \sin \theta_P  (m_K^2-m_{\pi }^2)
\nonumber\\
&&\;\;\;\; \; \; \; ~+\cos \theta_P ( 4 \cos2\theta_V-\sqrt{2} \sin2\theta_V )(4 m_K^2-m_{\pi }^2)~\bigg\}~ BW_{RR}[\eta,\phi,\rho,Q^2,s]~D_{R\eta 2}\nonumber\\
&&+ \frac{4F_V(1+8 \sqrt{2} \alpha_V\frac{2m_K^2-m_\pi^2}{M_V^2}) G_V}{3\sqrt{2} F^3 }\sin \delta \cos \theta _V
\nonumber\\
&&\;\;\;~\times \cos \delta  \cos \theta_P ( -4 \cos2\theta_V+\sqrt{2} \sin2\theta_V ) BW_{RR}[\eta,\phi,\omega,Q^2,s]~D_{R\eta 1}(Q^2,s,m_{\eta}^2)\nonumber\\
&&- \frac{4F_V(1+8 \sqrt{2} \alpha _V\frac{2m_K{}^2-m_\pi^2}{M_V^2}) G_V}{9\sqrt{2} F^3 }\sin \delta \cos \theta _V \cos \delta \nonumber\\
&&\;\;\;~\times \bigg\{ 4 ( 2 \sqrt{2} \cos2\theta_V-\sin2\theta_V ) \sin \theta_P (m_K^2-m_{\pi }^2)+\cos \theta_P ( 4 \cos2\theta_V
-\sqrt{2} \sin2\theta_V ) \nonumber\\
&&\;\;\;\;\; \; \; \; \;  ~\times (4m_K^2-m_{\pi }^2) \bigg\} ~BW_{RR}[\eta,\phi,\omega,Q^2,s]~D_{R\eta 2}.
\end{eqnarray}

\subsection{Notation}
\label{sub:3}
Notation employed in the previous Section and Appendix~\ref{app:B} is specified here:
\begin{eqnarray}
C_{R\pi}(Q^2,x)&=&(c_1-c_2+c_5) \, Q^2- (c_1-c_2-c_5+2 c_6) \,  x+ (c_1+c_2+8 c_3-c_5) \,  m_\pi^2 \; ,  \nonumber \\
D_{R\pi}(Q^2,x)&=& d_3 \, (Q^2+x)+(d_1+8d_2-d_3) \, m_\pi^2 \; ,  \nonumber \\
G_{R\pi}(Q^2)&=& (g_1+2g_2-g_3) \, (Q^2-3m_\pi^2)-3g_2\, (Q^2-3 m_\pi^2)+ 3 (2 g_4+g_5) \,m_\pi^2 \; , \nonumber \\
C_{R\eta1}(Q^2,x,m^2)&=&  (c_1-c_2+c_5) Q^2-(c_1-c_2-c_5+2 c_6) x+(c_1+c_2-c_5)m^2 \; , \nonumber \\
C_{R\eta2}&=& 8 \, c_3 \;  , \nonumber \\
D_{R\eta 1}(Q^2,x,m^2)&=&d_3(Q^2+x)+(d_1-d_3) \, m^2 \; , \nonumber \\
D_{R\eta 2}&=& 8 \, d_2 \; , \nonumber \\
G_{R\eta}(Q^2,s)&=&(g_1+2g_2-g_3) \, (s-2m_{\pi }^2)+g_2 \, (-Q^2+2m_{\pi }^2+m_{\eta }^2)+ (2g_4+g_5) \,  m_{\pi }^2 \; , \nonumber \\
BW[V,x]&=& \frac{1 }{M_{V }^2-i \Gamma_{V }(x) M_{V }-x} \, , \nonumber \\
BW_R[P,V,x]&=& BW[V,x]+ \beta_P' BW[V',x] + \beta_P'' BW[V'',x] \, , \nonumber \\
BW_{RR}[P,V_1,V_2,x,y] &=& BW_R[P,V_1,x] \, BW_R[P,V_2,y] \, ,
\end{eqnarray}
where the $c_i$, $d_i$ and $g_i$ coupling constants have been defined in Eqs.~(\ref{eq:lv4},\ref{eq:lvv2}). Moreover constraints on those couplings
(or appropriate combinations of them) have been obtained in Section~\ref{sec:4} and recalled in Eqs.~(\ref{eq:rescons11}-\ref{eq:rescons2}).

\section{Decay widths involving vector resonances}
\label{app:B}
We collect here the analytical expressions for the decay widths of vector resonances that have been employed in the fit analyses,
as explained in Section~\ref{sec:6}. The notation is the same as the one specified in Subsection~\ref{sub:3}.

\subsection{Two-body decays}
\begin{eqnarray}
\Gamma_{\rho \to \pi \pi }&=& \frac{G_V^2 \, M_{\rho}^3}{48 \pi F^4} \cos^2 \delta  \left(1 - \frac{4 m_{\pi}^2}{M_{\rho}^2}\right)^{3/2} , \nonumber \\
\Gamma_{\omega \to \pi \pi }&=&\frac{G_V^2 \, M_{\omega}^3}{48 \pi F^4} \sin^2 \delta  \left(1 - \frac{4 m_{\pi}^2}{M_{\omega}^2}\right)^{3/2} , \nonumber \\
\Gamma_{\phi \to \pi \pi }&=& \frac{ \alpha^2 \, \pi  \, F_V^2}{9 \, M_{\phi}} \left( 1+ 8 \sqrt{2} \alpha_V \frac{2 m_K^2-m_{\pi}^2}{M_V^2} \right)^2
 \cos^2 \theta_V \left(1 - \frac{4 m_{\pi}^2}{M_{\phi}^2}\right)^{3/2}, \nonumber  \\
\Gamma_{\rho \to \ell^+ \ell^-}&=& \frac{4 \, \alpha^2 \, \pi \, F_V^2}{3 \, M_{\rho}} \left( 1 + 8 \sqrt{2} \alpha_V \frac{m_{\pi}^2}{M_V^2} \right)^2 \,\left( \cos \delta +\frac{1}{\sqrt{3}}
\sin \theta _V \sin \delta \right)^2 \nonumber \\
&& \times \left( 1 + \frac{2 m_\ell^2}{M_{\rho}^2} \right) \left( 1 - \frac{4 m_\ell^2}{M_{\rho}^2} \right)^{1/2} , \\
\Gamma_{\omega \to \ell^+ \ell^-}&=&\frac{4 \, \alpha^2 \, \pi \, F_V^2}{27 \, M_{\omega}} \left( 1 + 8 \sqrt{2} \alpha_V \frac{m_{\pi}^2}{M_V^2} \right)^2 (\sqrt{3}\sin \theta _V\cos \delta -3\sin \delta )^2
\nonumber \\
&& \times  \left( 1 + \frac{2 m_\ell^2}{M_{\omega}^2} \right) \left( 1 - \frac{4 m_\ell^2}{M_{\omega}^2} \right)^{1/2} , \nonumber \\
\Gamma_{\phi \to \ell^+ \ell^-}&=&\frac{4 \,  \alpha^2 \, \pi \, F_V^2}{9 \, M_{\phi}} \left(1+8 \sqrt{2} \alpha _V\frac{2m_K^2-m_{\pi
}^2}{M_V^2}\right)^2 \cos^2 \theta_V \left( 1 + \frac{2 m_\ell^2}{M_{\phi}^2} \right) \left( 1 - \frac{4 m_\ell^2}{M_{\phi}^2} \right)^{1/2} , \nonumber
\end{eqnarray}
for $\ell = e$, $\mu$. Notice that $\omega \rightarrow \pi \pi$ is isospin violating and goes through $\rho-\omega$ mixing. The mixing between $\phi$ and $\rho$  is
essentially zero because the Okubo-Zweig-Iizuka rule, therefore in $\phi \rightarrow \pi \pi$  we only take into account the electromagnetic contribution.
\par
Radiative widths are given by:
\begin{eqnarray}
 \Gamma_{V \rightarrow P \gamma} & = & \frac{\alpha}{24} M_V \left(1-\frac{m_P^2}{M_V^2}\right)^3 |F_{V \rightarrow P \gamma}|^2 \, , \nonumber \\
\Gamma_{\eta' \rightarrow V \gamma} & = & \frac{\alpha}{8} \frac{m_{\eta'}^3}{M_V^2} \left(1-\frac{M_V^2}{m_{\eta'}^2}\right)^3 |F_{\eta' \rightarrow V \gamma}|^2 \, ,
\end{eqnarray}
where
\begin{eqnarray}
F_{\omega \to \pi^0 \gamma} &=&\frac{2\sqrt{2} }{3M_V F}C_{R\pi}(0, M_{\omega}^2 )
\left(\sqrt{3} \cos \delta (\sqrt{2}\cos \theta_V+\sin \theta_V )-\sin  \delta \right) \nonumber \\
&-&
\frac{4  F_V \left(1+8\sqrt{2}\alpha_V\frac{m_{\pi }^2}{M_V^2}\right)}{3 M_{\rho }^2 F} D_{R\pi}(0, M_{\omega }^2 )
( \sin \theta_V \sin \delta +\sqrt{3} \cos  \delta ) \cos 2\delta (\sqrt{2}\cos \theta_V+\sin \theta_V ) \nonumber\\
&+&
\frac{4 F_V \left(1+8\sqrt{2}\alpha_V\frac{m_{\pi }^2}{M_V^2}\right) }{3 M_{\omega}^2 F} D_{R\pi}(0, M_{\omega }^2)
(\sin \theta_V \cos \delta -\sqrt{3}\sin  \delta )\sin 2\delta (\sqrt{2}\cos  \theta_V +\sin \theta_V ) \nonumber \\
&-&
\frac{4 F_V \left(1+8\sqrt{2}\alpha_V\frac{2m_K^2-m_{\pi}^2}{M_V^2}\right)}{3M_{\phi}^2 F} D_{R\pi}(0, M_{\omega }^2)
\cos \theta_V \sin \delta (-\cos\theta_V+\sqrt{2}\sin \theta_V ) \, ,
\end{eqnarray}
\begin{eqnarray}
F_{\omega \to \eta \gamma}&=&\frac{2\sqrt{2} }{3 M_V  F} C_{R\eta 1}(0, M_{\omega }^2,m_{\eta}^2) \nonumber \\
&& \left\{ \sqrt{3} \sin \delta (-\cos\theta _P+\sqrt{2} \sin \theta _P )+\cos \delta
[ \sqrt{2} \cos \theta _V\cos \theta _P-\sin \theta _V (\cos \theta _P+\sqrt{2} \sin \theta _P) ]\right\} \nonumber \\
&+&
\frac{2\sqrt{2} }{9 M_V  F}C_{R\eta 2}
\left\{ 4 \cos \delta  \left(\sqrt{2} \cos( \theta_V +\theta _P )-2 \cos  \theta _P \sin\theta_V+\cos \theta _V\sin  \theta _P\right) m_K^2\right. \nonumber \\
&&-\left(3 \sqrt{3} \sin \delta  (\cos \theta _P-\sqrt{2} \sin \theta _P)\right. \nonumber \\
&&\left.\left.+\cos \delta[\sqrt{2} \cos(\theta_V +\theta _P)-5 \cos  \theta_P \sin\theta_V+4 \cos  \theta _V\sin  \theta _P]\right) m_{\pi}^2\right\} \nonumber \\
&-&
\frac{ F_V\left(1+8\sqrt{2}\alpha_V\frac{m_{\pi }^2}{M_V^2}\right)}{3\sqrt{2}M_{\rho }^2 F}D_{R\eta1}(0,M_{\omega }^2,m_{\eta}^2) \nonumber \\
&&
(\sin\theta_V\sin  \delta +\sqrt{3}\cos  \delta ) \cos \theta_P \sin2\delta (-3 \sqrt{2}+\sqrt{2} \cos2\theta_V+4 \sin2\theta_V) \nonumber \\
&-&
\frac{ F_V\left(1+8\sqrt{2}\alpha_V\frac{m_{\pi }^2}{M_V^2}\right) }{9\sqrt{2} M_{\rho }^2 F} D_{R\eta2} \nonumber\\
&&
(\sin\theta_V\sin  \delta +\sqrt{3}\cos  \delta )\sin2\delta
 \left\{4 (-3+\cos2\theta_V+2 \sqrt{2} \sin2\theta_V ) \sin  \theta _P (m_K^2-m_{\pi }^2)\right. \nonumber \\
&&\left.+\cos  \theta _P (-3\sqrt{2}+\sqrt{2} \cos2\theta_V+4 \sin2\theta_V) ( 4 m_K^2-m_{\pi }^2 )\right\} \nonumber \\
&-&
\frac{2\sqrt{2} F_V\left(1+8\sqrt{2}\alpha_V\frac{m_{\pi }^2}{M_V^2}\right) }{3 M_{\omega }^2 F}D_{R\eta 1}(0,M_{\omega }^2,m_{\eta}^2)
(\sin\theta_V \cos \delta -\sqrt{3}\sin  \delta ) \nonumber \\
&&\left\{\cos \theta_P [\sqrt{2} \sin^2 \delta+\cos^2 \delta \sin \theta_V (4 \cos\theta_V -\sqrt{2}\sin\theta_V )]-2 \sin \theta_P\right\} \nonumber \\
&-&
\frac{ F_V\left(1+8\sqrt{2}\alpha_V\frac{m_{\pi }^2}{M_V^2}\right) }{ 9\sqrt{2} M_{\omega }^2  F} D_{R\eta2}
(\sin\theta_V \cos \delta -\sqrt{3}\sin \delta ) \nonumber \\
& &
\left\{8 \cos^2\delta \left(\cos  \theta _P (-3 \sqrt{2}+\sqrt{2} \cos2\theta_V+4 \sin2\theta_V )\right.\right. \nonumber \\
&&\left.\left.+( -3+\cos2 \theta_V+2 \sqrt{2} \sin2\theta_V ) \sin  \theta _P\right) m_K^2\right.
+\left(12 \sin^2\delta ( \sqrt{2} \cos\theta_P- 2 \sin\theta_P )\right. \nonumber \\
&&\left.+\cos^2\delta [-9 \sqrt{2}\cos(2\theta_V -\theta _P)+18 \sqrt{2} \cos \theta_P \right. \nonumber \\
&&\left.\left.+7 \sqrt{2}\cos(2 \theta_V +\theta _P)-8 \sin(2 \theta_V +\theta _P) ]\right)m_{\pi }^2\right\} \nonumber\\
&+&
\frac{\sqrt{2} F_V \left(1+8\sqrt{2}\alpha_V\frac{2m_K^2-m_{\pi }^2}{M_V^2}\right)}{3 M_{\phi }^2 F}
D_{R\eta 1}(0,M_{\omega }^2,m_{\eta}^2) \nonumber \\
&&\left\{\cos\theta_V \cos \delta  \cos \theta_P ( -4 \cos2\theta_V +\sqrt{2} \sin2\theta_V )\right\} \nonumber \\
&-&
\frac{\sqrt{2} F_V \left(1+8\sqrt{2}\alpha_V\frac{2m_K^2-m_{\pi }^2}{M_V^2}\right)}{9 M_{\phi}^2 F}D_{R\eta 2}
\cos\theta_V\cos \delta \nonumber \\
&&\left\{ 4 (2 \sqrt{2} \cos 2\theta_V-\sin2\theta_V ) \sin  \theta _P (m_K^2-m_{\pi }^2)\right. \nonumber \\
&&\left.+\cos  \theta _P (4 \cos2\theta_V-\sqrt{2} \sin2\theta_V ) (4 m_K^2-m_{\pi }^2)\right\} \, ,
\end{eqnarray}
\begin{eqnarray}
F_{\rho^0 \to \pi^0 \gamma} &=&\frac{2\sqrt{2} }{3 M_V  F}C_{R\pi}(0,M_{\rho }^2)\left(\cos \delta +\sqrt{6}\cos  \theta_V \sin
\delta +\sqrt{3}\sin \delta \sin \theta_V\right)\nonumber \\
&-&
\frac{4 F_V\left(1+8\sqrt{2}\alpha_V\frac{m_{\pi }^2}{M_V^2}\right) }{3 M_{\rho }^2  F}D_{R\pi}(0,M_{\rho}^2)
(\sin \theta_V\sin \delta +\sqrt{3}\cos  \delta )\sin 2\delta (\sqrt{2}\cos \theta_V+\sin \theta_V)\nonumber \\
&-&
\frac{4 F_V\left(1+8\sqrt{2}\alpha_V\frac{m_{\pi }^2}{M_V^2}\right) }{3 M_{\omega }^2  F}D_{R\pi}(0,M_{\rho}^2)
(\sin \theta_V \cos \delta -\sqrt{3}\sin  \delta )\cos  2\delta (\sqrt{2}\cos\theta_V+\sin\theta_V )\nonumber \\
&-&\frac{4 F_V \left(1+8\sqrt{2}\alpha_V\frac{2m_K^2-m_{\pi }^2}{M_V^2}\right)}{3 M_{\phi }^2 F}D_{R\pi}(0,M_{\rho }^2)
\cos \theta_V \cos \delta (\cos\theta_V-\sqrt{2}\sin\theta_V) \, ,
\end{eqnarray}
\begin{eqnarray}
F_{\rho^\pm \to \pi^\pm \gamma} &=&\frac{2\sqrt{2} }{3 M_V  F} C_{R\pi} (0, M_{\rho}^2)\nonumber \\
&-&
\frac{4 F_V\left(1+8\sqrt{2}\alpha_V\frac{m_{\pi }^2}{M_V^2}\right) }{3 M_{\rho }^2 F}D_{R\pi}(0,M_{\rho }^2)
\sin \delta (\sqrt{2}\cos \theta_V+\sin \theta_V )(\sin \theta_V \sin \delta +\sqrt{3}\cos \delta )\nonumber \\
&-&
\frac{4 F_V\left(1+8\sqrt{2}\alpha_V\frac{m_{\pi }^2}{M_V^2}\right) }{3 M_{\omega}^2  F}D_{R\pi}(0,M_{\rho }^2)
\cos \delta (\sqrt{2}\cos \theta_V+\sin \theta_V )(\sin \theta_V \cos \delta -\sqrt{3}\sin \delta )\nonumber \\
&-&
\frac{4F_V \left(1+8\sqrt{2}\alpha_V\frac{2m_K^2-m_{\pi }^2}{M_V^2}\right)}{3 M_{\phi }^2 F}D_{R\pi}(0,M_{\rho}^2)
\cos \theta_V(\cos \theta_V-\sqrt{2}\sin \theta_V) \, ,
\end{eqnarray}
\begin{eqnarray}
F_{\rho^0 \to \eta \gamma} &=&\frac{2\sqrt{2} }{3 M_V  F} C_{R\eta 1}(0,M_{\rho }^2,m_\eta^2) \left\{\sqrt{3} \cos  \delta
(\cos \theta _P-\sqrt{2} \sin  \theta _P)\right.\nonumber \\
&&\left.+\sin \delta  [\sqrt{2} \cos  \theta_V\cos \theta_P-\sin \theta_V (\cos \theta _P+\sqrt{2}\sin \theta _P)]\right\}\nonumber \\
&+&
\frac{2\sqrt{2} }{9 M_V  F}C_{R\eta 2}
\left\{4 \sin \delta \left(\sqrt{2} \cos(\theta_V +\theta _P)-2 \cos \theta _P \sin\theta_V
+\cos \theta _V\sin \theta _P\right) m_K^2\right.\nonumber \\
&&+\left(3 \sqrt{3} \cos  \delta  (\cos  \theta _P-\sqrt{2} \sin  \theta _P)\right.\nonumber \\
&&\left.\left.-\sin\delta [\sqrt{2} \cos(\theta_V +\theta _P)-5 \cos  \theta _P \sin\theta_V +4 \cos  \theta _V\sin  \theta _P]\right) m_{\pi}^2\right\}\nonumber \\
&-&
\frac{2\sqrt{2} F_V\left(1+8\sqrt{2}\alpha_V\frac{m_{\pi }^2}{M_V^2}\right) }{3 M_{\rho }^2 F}D_{R\eta 1}(0,M_{\rho }^2,m_{\eta}^2)
(\sin\theta_V\sin  \delta +\sqrt{3}\cos  \delta )\nonumber \\
&&\left\{\cos^2\delta (\sqrt{2}\cos  \theta _P-2 \sin  \theta _P)
+\sin^2\delta [\cos  \theta _P \sin  \theta _V(4 \cos\theta_V-\sqrt{2} \sin\theta_V)-2 \sin  \theta _P]\right\}\nonumber \\
&-&
\frac{ F_V\left(1+8\sqrt{2}\alpha_V\frac{m_{\pi }^2}{M_V^2}\right) } {9\sqrt{2}M_{\rho }^2 F} D_{R\eta2}
(\sin\theta_V\sin \delta +\sqrt{3}\cos \delta )\nonumber \\
&&\left\{8 \sin^2\delta \left( \cos  \theta _P (-3 \sqrt{2}+\sqrt{2} \cos 2\theta_V+4 \sin2\theta_V )\right.\right.\nonumber \\
&&\left.+(-3+\cos 2\theta_V+2 \sqrt{2} \sin 2\theta_V)\sin  \theta _P \right) m_K^2+\left(12 \cos^2\delta (\sqrt{2} \cos  \theta _P-2 \sin  \theta _P )\right.\nonumber \\
&&\left.+\sin^2\delta [-9 \sqrt{2} \cos(2
\theta_V -\theta _P)+18 \sqrt{2} \cos  \theta _P\right.\nonumber \\
&&\left.\left.+7 \sqrt{2} \cos(2 \theta_V +\theta _P)-8 \sin(2 \theta_V +\theta _P)]\right)m_{\pi }^2\right\}\nonumber \\
&-&
\frac{ F_V\left(1+8\sqrt{2}\alpha_V\frac{m_{\pi}^2}{M_V^2}\right) }{3\sqrt{2} M_{\omega }^2 F}D_{R\eta 1}(0,M_{\rho }^2,m_{\eta}^2)
(\sin\theta_V\cos  \delta -\sqrt{3}\sin  \delta )\nonumber \\
&&\left\{\cos \theta _P \sin2\delta (-3\sqrt{2}+\sqrt{2} \cos 2\theta_V+4 \sin 2\theta_V)\right\}\nonumber \\
&-&
\frac{F_V\left(1+8\sqrt{2}\alpha_V\frac{m_{\pi }^2}{M_V^2}\right) }{9\sqrt{2} M_{\omega }^2  F}D_{R\eta 2}
(\sin\theta_V\cos  \delta -\sqrt{3}\sin  \delta )\sin2\delta\nonumber \\
&&\left\{4 (-3+\cos2\theta_V+2 \sqrt{2} \sin2\theta_V) \sin \theta _P (m_K^2-m_{\pi }^2)\right.\nonumber \\
&&\left.+\cos  \theta _P (-3\sqrt{2}+\sqrt{2} \cos2\theta_V+4 \sin 2\theta_V ) (4 m_K^2-m_{\pi}^2)\right\}\nonumber \\
&+&
\frac{2 F_V\left(1+8\sqrt{2}\alpha_V\frac{2m_K^2-m_{\pi }^2}{M_V^2}\right)}{3M_{\phi }^2 F}D_{R\eta 1}(0,M_{\rho }^2,m_{\eta}^2)
\cos \theta_V \cos \theta_P \sin \delta  (-4 \cos2\theta_V+\sqrt{2} \sin2\theta_V)\nonumber \\
&-&
\frac{\sqrt{2} F_V \left(1+8\sqrt{2}\alpha_V\frac{2m_K^2-m_{\pi }^2}{M_V^2}\right)}{9 M_{\phi}^2 F}D_{R\eta 2}
\cos\theta_V \sin \delta \nonumber \\
&&\left\{4 (2 \sqrt{2} \cos2\theta_V-\sin2\theta_V ) \sin  \theta _P (m_K^2-m_{\pi }^2)\right.\nonumber \\
&&\left.+\cos  \theta _P (4 \cos2\theta_V-\sqrt{2} \sin2\theta_V) (4 m_K^2-m_{\pi }^2)\right\} \, ,
\end{eqnarray}
\begin{eqnarray}
F_{\phi \to \pi^0 \gamma} &=&\frac{2\sqrt{6} }{3 M_V  F} C_{R\pi}(0,M_{\phi }^2)(\cos \theta_V-\sqrt{2}\sin\theta_V) \\
&-&
\frac{4F_V\left(1+8\sqrt{2}\alpha_V\frac{m_{\pi }^2}{M_V^2}\right) }{3 M_{\rho }^2 F}D_{R\pi}(0,M_{\phi }^2)
(\sin\theta_V\sin \delta +\sqrt{3}\cos \delta )\cos \delta (\cos \theta_V-\sqrt{2}\sin \theta_V)\nonumber \\
&-&
\frac{4 F_V\left(1+8\sqrt{2}\alpha_V\frac{m_{\pi }^2}{M_V^2}\right) }{3 M_{\omega }^2 F} D_{R\pi}(0,M_{\phi }^2)
(\sin\theta_V\cos  \delta -\sqrt{3}\sin \delta )\sin \delta (-\cos \theta_V+\sqrt{2}\sin \theta_V) \,  \nonumber ,
\end{eqnarray}
\begin{eqnarray}
F_{\phi \to \eta \gamma} &=&\frac{2\sqrt{2} }{3M_V  F}C_{R\eta 1}(0,M_{\phi}^2,m_{\eta}^2)
\left\{-\sqrt{2} \cos  \theta _P \sin\theta_V-\cos  \theta _V( \cos \theta _P+\sqrt{2} \sin \theta _P)\right\}\nonumber \\
&+&
\frac{\sqrt{2} }{9 M_V  F}C_{R\eta 2}
 \left\{-4 \left( 3 \cos(\theta_V -\theta _P)+\cos(\theta_V +\theta _P)+2 \sqrt{2}\sin(\theta_V +\theta _P) \right) m_K^2 \right.\nonumber \\
&&\left.+\left(9 \cos(\theta_V -\theta _P)+\cos(\theta_V +\theta _P)+2\sqrt{2} \sin(\theta_V +\theta _P) \right) m_{\pi }^2\right\}\nonumber \\
&+&
\frac{\sqrt{2} F_V \left(1+8\sqrt{2}\alpha_V\frac{m_{\pi}^2}{M_V^2}\right)}{3 M_{\rho }^2 F}D_{R\eta 1}(0,M_{\phi }^2,m_{\eta}^2)
(\sin\theta_V\sin  \delta +\sqrt{3}\cos  \delta )\nonumber \\
&&\cos  \theta _P \sin  \delta  (-4\cos 2\theta_V+\sqrt{2} \sin 2\theta_V )\nonumber \\
&-&
\frac{\sqrt{2} F_V\left(1+8\sqrt{2}\alpha_V\frac{m_{\pi }^2}{M_V^2}\right)}{9 M_{\rho }^2 F}D_{R\eta 2}
(\sin\theta_V\sin  \delta +\sqrt{3}\cos \delta)\sin \delta\nonumber \\
&& \left\{4(2 \sqrt{2} \cos 2\theta_V-\sin 2\theta_V) \sin  \theta _P (m_K^2-m_{\pi }^2)\right.\nonumber \\
&&\left.+\cos  \theta _P (4\cos 2\theta_V-\sqrt{2} \sin 2\theta_V)(4 m_K^2-m_{\pi }^2)\right\}\nonumber \\
&+&
\frac{\sqrt{2}F_V\left(1+8\sqrt{2}\alpha_V\frac{m_{\pi}^2}{M_V^2}\right)}{3 M_{\omega }^2 F}D_{R\eta 1}(0,M_{\phi }^2,m_{\eta}^2)
(\sin\theta_V\cos  \delta -\sqrt{3}\sin \delta )\nonumber \\
&&\cos \delta  \cos \theta _P (-4\cos 2\theta_V+\sqrt{2} \sin 2\theta_V)\nonumber \\
&-&
\frac{\sqrt{2} F_V\left(1+8\sqrt{2}\alpha_V\frac{m_{\pi}^2}{M_V^2}\right)}{9 M_{\omega }^2 F}D_{R\eta 2}
(\sin\theta_V\cos  \delta -\sqrt{3}\sin  \delta )\cos  \delta  \nonumber \\
&&\left\{4 (2 \sqrt{2} \cos 2\theta_V-\sin 2\theta_V) \sin  \theta _P (m_K^2-m_{\pi }^2)\right.\nonumber \\
&&\left.+\cos  \theta _P (4 \cos 2\theta_V-\sqrt{2} \sin 2\theta_V ) (4 m_K^2-m_{\pi }^2)\right\}\nonumber \\
&-&
\frac{2\sqrt{2} F_V \left(1+8\sqrt{2}\alpha_V\frac{2m_K^2-m_{\pi }^2}{M_V^2}\right)}{3 M_{\phi}^2 F}D_{R\eta 1}(0,M_{\phi }^2,m_{\eta}^2)
\cos\theta_V\nonumber \\
&&\left\{-\cos  \theta _V\cos  \theta _P (\sqrt{2} \cos\theta_V+4 \sin\theta_V)-2 \sin \theta _P\right\}\nonumber \\
&-&
\frac{\sqrt{2} F_V\left(1+8\sqrt{2}\alpha_V\frac{2m_K^2-m_{\pi }^2}{M_V^2}\right)}{9 M_{\phi }^2 F}D_{R\eta 2}\cos\theta_V\nonumber \\
&&\left\{ (\sqrt{2} \cos\theta_V-2 \sin\theta_V)^2 (\sqrt{2}\cos  \theta _P-2 \sin  \theta _P) m_{\pi }^2\right.\nonumber \\
&&\left.-4 (\sqrt{2} \cos\theta_V+\sin\theta_V)^2 (\sqrt{2} \cos \theta_P+\sin \theta _P) (2 m_K^2-m_{\pi}^2) \right\} \, ,
\end{eqnarray}
\begin{eqnarray}
F_{\eta' \to \omega \gamma} &=&\frac{2\sqrt{2} }{3M_V  F}C_{R\eta 1}(0,M_{\omega }^2,m_{\eta'}^2)
\left\{\cos  \delta  \sin  \theta_V(\sqrt{2} \cos  \theta _P-\sin  \theta _P)\right.\nonumber \\
&&\left.+\sqrt{2} \cos  \delta  \cos  \theta _V\sin  \theta _P-\sqrt{3} \sin  \delta  (\sqrt{2}
\cos  \theta _P+\sin  \theta _P)\right\}\nonumber \\
&+&
\frac{\sqrt{2} }{9 M_V  F} C_{R\eta 2} \left\{4 \cos  \delta  \left(-3 \cos(\theta_V -\theta _P)+\cos(\theta_V +\theta _P)+2
\sqrt{2} \sin(\theta_V +\theta _P) \right) m_K^2\right.\nonumber \\
&&+\left(-6 \sqrt{3} \sin  \delta  (\sqrt{2} \cos  \theta _P+\sin  \theta _P)\right.\nonumber \\
&&\left.\left.-\cos \delta  [-9 \cos(\theta_V -\theta _P)+\cos(\theta_V +\theta _P)+2 \sqrt{2} \sin(\theta_V +\theta _P)]\right)
m_{\pi }^2\right\}\nonumber \\
&-&
\frac{ F_V\left(1+8\sqrt{2}\alpha_V\frac{m_{\pi}^2}{M_V^2}\right) }{3\sqrt{2} M_{\rho }^2 F}D_{R\eta 1}(0,M_{\omega }^2,m_{\eta'}^2)\nonumber \\
&&\left\{(\sin\theta_V\sin  \delta +\sqrt{3}\cos  \delta )\sin2\delta(-3 \sqrt{2}+\sqrt{2}\cos 2\theta_V+4 \sin 2\theta_V) \sin \theta_P\right\}\nonumber \\
&-&
\frac{ F_V\left(1+8\sqrt{2}\alpha_V\frac{m_{\pi }^2}{M_V^2}\right) }{9\sqrt{2} M_{\rho }^2 F}D_{R\eta 2}
(\sin\theta_V\sin  \delta +\sqrt{3}\cos  \delta )\sin2\delta \nonumber \\
&&\left\{-4 \cos  \theta _P (-3+\cos 2\theta_V+2 \sqrt{2} \sin 2\theta_V)(m_K^2-m_{\pi }^2)\right.\nonumber \\
&&\left.+(-3 \sqrt{2}+\sqrt{2}\cos 2\theta_V+4 \sin 2\theta_V) \sin  \theta _P (4 m_K^2-m_{\pi }^2)\right\}\nonumber \\
&-&
\frac{2\sqrt{2} F_V\left(1+8\sqrt{2}\alpha_V\frac{m_{\pi }^2}{M_V^2}\right) }{3 M_{\omega }^2 F}D_{R\eta 1}(0,M_{\omega }^2,m_{\eta'}^2)
(\sin\theta_V\cos  \delta -\sqrt{3}\sin  \delta )\nonumber \\
&&\left\{\sin^2\delta (2\cos  \theta _P+\sqrt{2} \sin  \theta _P)
+\cos^2\delta [ 2 \cos  \theta _P+\sin  \theta _V(4 \cos\theta_V-\sqrt{2} \sin\theta_V) \sin  \theta _P ] \right\}\nonumber \\
&-&
\frac{\sqrt{2}F_V\left(1+8\sqrt{2}\alpha_V\frac{m_{\pi }^2}{M_V^2}\right) }{9 M_{\omega }^2 F} D_{R\eta 2}
(\sin\theta_V\cos  \delta -\sqrt{3}\sin  \delta )\nonumber \\
&&\left\{-4 \cos^2\delta \left(\cos  \theta _P (-3+\cos 2\theta_V+2 \sqrt{2} \sin 2\theta_V)\right.\right.\nonumber \\
&&\left.-(-3 \sqrt{2}+\sqrt{2}\cos 2\theta_V+4 \sin 2\theta_V) \sin  \theta _P\right) m_K^2
+\left(6 \sin^2\delta (2 \cos  \theta _P+\sqrt{2} \sin  \theta _P)\right.\nonumber \\
&&\left.\left.+\cos^2\delta \left(4 \cos(2 \theta_V +\theta_P)
+\sqrt{2} [8 \cos  \theta _P \sin 2\theta_V-(-9+\cos 2\theta_V) \sin \theta _P]\right)\right) m_{\pi }^2\right\}\nonumber \\
&+&
\frac{\sqrt{2} F_V \left(1+8\sqrt{2}\alpha_V\frac{2m_K^2-m_{\pi }^2}{M_V^2}\right)}{3 M_{\phi}^2 F}D_{R\eta 1}(0,M_{\omega }^2,m_{\eta'}^2)\nonumber \\
&&\left\{\cos\theta_V \cos \delta  (-4 \cos 2\theta_V+\sqrt{2} \sin 2\theta_V) \sin \theta _P\right\}\nonumber \\
&-&
\frac{\sqrt{2} F_V \left(1+8\sqrt{2}\alpha_V\frac{2m_K^2-m_{\pi }^2}{M_V^2}\right)}{9M_{\phi}^2 F} D_{R\eta 2}
\cos\theta_V \cos \delta  \nonumber \\
&&\left\{-4 \cos  \theta _P (2 \sqrt{2}\cos 2\theta_V-\sin 2\theta_V) (m_K^2-m_{\pi }^2)\right.\nonumber \\
&&\left.+(4 \cos 2\theta_V-\sqrt{2} \sin 2\theta_V)\sin  \theta _P (4 m_K^2-m_{\pi }^2)\right\} \, ,
\end{eqnarray}
\begin{eqnarray}
F_{\eta' \to \rho \gamma} &=&\frac{2\sqrt{2} }{3 M_V  F}C_{R\eta 1}(0,M_{\rho }^2,m_{\eta'}^2)
 \left\{\sqrt{3} \cos  \delta  (\sqrt{2}\cos \theta _P+\sin  \theta _P)\right.\nonumber \\
&&\left.+\sin  \delta  [ \sqrt{2} \cos  \theta _P \sin\theta_V+(\sqrt{2} \cos\theta_V-\sin\theta_V) \sin  \theta _P ] \right\}\nonumber \\
&+&
\frac{\sqrt{2} }{9 M_V  F}C_{R\eta 2}
\left\{4 \sin  \delta  \left(-3 \cos(\theta_V -\theta _P)+\cos(\theta_V +\theta _P)+2 \sqrt{2} \sin(\theta_V
+\theta _P)\right) m_K^2\right.\nonumber \\
&&+\left(6 \sqrt{3} \cos \delta  (\sqrt{2} \cos  \theta _P+\sin  \theta _P)\right.\nonumber \\
&&\left.\left.-\sin  \delta  [-9 \cos(\theta_V-\theta _P)+\cos(\theta_V +\theta _P)+2 \sqrt{2} \sin(\theta_V +\theta _P)]\right) m_{\pi }^2\right\}\nonumber \\
&-&
\frac{2\sqrt{2} F_V\left(1+8\sqrt{2}\alpha_V\frac{m_{\pi }^2}{M_V^2}\right)}{3 M_{\rho}^2 F}D_{R\eta 1}(0,M_{\rho }^2,m_{\eta'}^2)
(\sin\theta_V\sin  \delta +\sqrt{3}\cos  \delta )\nonumber \\
&&\left\{\cos^2\delta (2 \cos  \theta_P+\sqrt{2} \sin  \theta _P)
+\sin^2\delta [2 \cos  \theta _P+\sin  \theta _V(4 \cos\theta_V-\sqrt{2} \sin\theta_V) \sin  \theta _P]\right\}\nonumber \\
&-&
\frac{\sqrt{2} F_V\left(1+8\sqrt{2}\alpha_V\frac{m_{\pi }^2}{M_V^2}\right)}{9 M_{\rho }^2 F} D_{R\eta 2}
(\sin\theta_V\sin \delta +\sqrt{3}\cos \delta )\nonumber \\
&&
\left\{-4 \sin^2\delta \left(\cos  \theta _P (-3+\cos 2\theta_V+2 \sqrt{2} \sin 2\theta_V)\right.\right.\nonumber \\
&&\left.-(-3 \sqrt{2}+\sqrt{2}
\cos 2\theta_V+4 \sin 2\theta_V) \sin  \theta _P\right) m_K^2+\left(6 \cos^2\delta (2 \cos  \theta _P+\sqrt{2} \sin  \theta _P)\right.\nonumber \\
&&\left.\left.+\sin^2\delta \left(4 \cos (2 \theta_V +\theta_P)
+\sqrt{2} [8 \cos  \theta _P \sin 2\theta_V-(-9+\cos 2\theta_V) \sin \theta_P]\right) \right) m_{\pi }^2\right\}\nonumber \\
&-&
\frac{\sqrt{2} F_V \left(1+8\sqrt{2}\alpha_V\frac{m_{\pi}^2}{M_V^2}\right)}{6 M_{\omega }^2 F}D_{R\eta 1}(0,M_{\rho }^2,m_{\eta'}^2)
(\sin\theta_V\cos  \delta -\sqrt{3}\sin  \delta )\nonumber \\
&&\left\{\sin2\delta (-3 \sqrt{2}+\sqrt{2} \cos 2\theta_V+4 \sin 2\theta_V) \sin  \theta _P\right\}\nonumber \\
&-&
\frac{\sqrt{2} F_V \left(1+8\sqrt{2}\alpha_V\frac{m_{\pi }^2}{M_V^2}\right)}{18 M_{\omega }^2 F}D_{R\eta 2}
(\sin\theta_V\cos  \delta -\sqrt{3}\sin  \delta )\sin2\delta \nonumber \\
&& \left\{-4 \cos \theta_P (-3+\cos 2\theta_V+2 \sqrt{2} \sin 2\theta_V) (m_K^2-m_{\pi }^2)\right.\nonumber \\
&&\left.+(-3 \sqrt{2}+\sqrt{2} \cos 2\theta_V+4 \sin 2\theta_V) \sin \theta _P (4 m_K^2-m_{\pi }^2)\right\}\nonumber \\
&+&
\frac{\sqrt{2} F_V\left(1+8\sqrt{2}\alpha_V\frac{2m_K^2-m_{\pi }^2}{M_V^2}\right)}{3M_{\phi
}^2 F}D_{R\eta 1}(0,M_{\rho }^2,m_{\eta'}^2)\nonumber \\
&&\left\{\cos\theta_V \sin \delta (-4 \cos 2\theta_V+\sqrt{2} \sin 2\theta_V) \sin \theta_P\right\}\nonumber \\
&-&
\frac{\sqrt{2} F_V\left(1+8\sqrt{2}\alpha_V\frac{2m_K^2-m_{\pi }^2}{M_V^2}\right)}{9M_{\phi}^2 F}D_{R\eta 2}\cos\theta_V\sin \delta \nonumber \\
&& \left\{-4 \cos  \theta _P (2 \sqrt{2} \cos 2\theta_V-\sin 2\theta_V) (m_K^2-m_{\pi }^2)\right.\nonumber \\
&&\left.+(4 \cos 2\theta_V-\sqrt{2} \sin 2\theta_V) \sin  \theta _P (4 m_K^2-m_{\pi}^2)\right\} \, ,
\end{eqnarray}
\begin{eqnarray}
F_{\phi \to \eta' \gamma} &=&\frac{2\sqrt{2} }{3 M_V  F} C_{R\eta 1}(0, M_{\phi }^2,m_{\eta'}^2)
\left\{\sqrt{2} \cos(\theta_V+\theta _P)-\cos  \theta _V\sin  \theta _P\right\}\nonumber \\
&+&
\frac{\sqrt{2} }{9 M_V F}C_{R\eta 2}
\left\{8 \left(\sqrt{2} \cos(\theta_V +\theta _P)+\cos  \theta _P \sin\theta_V-2 \cos  \theta _V\sin  \theta_P\right) m_K^2\right.\nonumber \\
&&\left.+\left(-2 \sqrt{2} \cos(\theta_V +\theta _P)-9 \sin(\theta_V -\theta _P)+\sin(\theta_V +\theta_P)\right) m_{\pi }^2\right\}\nonumber \\
&+&
\frac{\sqrt{2} F_V \left(1+8\sqrt{2}\alpha_V\frac{m_{\pi}^2}{M_V^2}\right)}{3M_{\rho }^2 F} D_{R\eta 1}(0,M_{\phi }^2,m_{\eta'}^2)
(\sin\theta_V\sin  \delta +\sqrt{3}\cos  \delta )\nonumber \\
&&\left\{\sin \delta (-4 \cos 2\theta_V+\sqrt{2}\sin 2\theta_V) \sin  \theta_P\right\}\nonumber \\
&-&
\frac{\sqrt{2} F_V \left(1+8\sqrt{2}\alpha_V\frac{m_{\pi }^2}{M_V^2}\right)}{9 M_{\rho }^2  F} D_{R\eta 2}
(\sin\theta_V\sin  \delta +\sqrt{3}\cos  \delta ) \sin \delta \nonumber \\
&& \left\{-4 \cos  \theta _P (2\sqrt{2} \cos 2\theta_V-\sin 2\theta_V)(m_K^2-m_{\pi }^2)\right.\nonumber \\
&&\left.+(4 \cos 2\theta_V-\sqrt{2} \sin 2\theta_V) \sin  \theta _P (4 m_K^2-m_{\pi }^2)\right\}\nonumber \\
&+&
\frac{\sqrt{2} F_V \left(1+8\sqrt{2}\alpha_V\frac{m_{\pi}^2}{M_V^2}\right)}{3 M_{\omega }^2 F}D_{R\eta 1}(0,M_{\phi }^2,m_{\eta'}^2)
(\sin\theta_V\cos \delta -\sqrt{3}\sin \delta )\nonumber \\
&&\left\{\cos \delta  (-4 \cos 2\theta_V+\sqrt{2}\sin 2\theta_V) \sin \theta _P\right\}\nonumber \\
&-&
\frac{\sqrt{2} F_V \left(1+8\sqrt{2}\alpha_V\frac{m_{\pi }^2}{M_V^2}\right)}{9 M_{\omega }^2 F} D_{R\eta 2}
(\sin\theta_V\cos \delta -\sqrt{3}\sin \delta ) \cos \delta\nonumber \\
&&\left\{-4 \cos  \theta _P (2\sqrt{2} \cos 2\theta_V-\sin 2\theta_V) (m_K^2-m_{\pi }^2)\right.\nonumber \\
&&\left.+(4 \cos 2\theta_V-\sqrt{2} \sin 2\theta_V) \sin  \theta _P (4 m_K^2-m_{\pi }^2)\right\}\nonumber \\
&-&
\frac{2\sqrt{2} F_V \left(1+8\sqrt{2}\alpha_V\frac{2m_K^2-m_{\pi }^2}{M_V^2}\right)}{3 M_{\phi}^2 F}D_{R\eta 1}(0,M_{\phi }^2,m_{\eta'}^2)
\cos\theta_V\nonumber \\
&&\left\{2 \cos  \theta _P-\cos  \theta _V(\sqrt{2} \cos\theta_V+4 \sin\theta_V) \sin  \theta _P\right\}\nonumber \\
&-&
\frac{\sqrt{2} F_V \left(1+8\sqrt{2}\alpha_V\frac{2m_K^2-m_{\pi }^2}{M_V^2}\right)}{9 M_{\phi }^2 F}D_{R\eta 2}
\cos\theta_V \nonumber \\
&&\left\{ (\sqrt{2} \cos\theta_V-2 \sin\theta_V)^2 (2 \cos \theta _P+\sqrt{2} \sin  \theta _P) m_{\pi }^2\right.\nonumber \\
&&\left.-4 (\sqrt{2} \cos\theta_V+\sin\theta_V)^2 (-\cos  \theta _P+\sqrt{2}\sin  \theta _P)(2 m_K^2-m_{\pi }^2)\right\} \, .
\end{eqnarray}

\subsection{Three-body decays}
The three pion decays of the vector resonances are given by:
\begin{equation} \label{eq:wp3pi}
\Gamma ( V \rightarrow \pi^+(p_1) \pi^-(p_2) \pi^0(p_3)) = \frac{1}{256 \, \pi^3 \, M_V^3} \int_{s_-}^{s_+} ds  \int_{t_-}^{t_+} dt \, {\cal P}(s,t) | \Omega_V |^2 \, ,
\end{equation}
for $V= \omega, \phi, \rho$, where $s=(p_1+p_2)^2$, $t=(p_1+p_3)^2$ and
\begin{equation} \label{eq:pw3pi}
 {\cal P}(s,t) = \frac{1}{12} \left[ (3 m_{\pi}^2+M_V^2-s) s t - st^2-m_{\pi}^2 ( m_{\pi}^2-M_V^2)^2 \right] \,,
\end{equation}
and the integration limits are:
\begin{eqnarray} \label{eq:intlim}
 s_+ &=& (M_V-m_{\pi})^2 \, , \nonumber \\
s_- &=& 4 m_{\pi}^2 \, , \nonumber \\
t_{\mp} &=& \frac{1}{4s} \left[ \left( M_V^2-m_{\pi}^2 \right)^2- \left( \lambda^{1/2}(s,m_{\pi}^2,m_{\pi}^2) \pm \lambda^{1/2}(M_V^2,s,m_{\pi}^2) \right)^2 \right]  .
\end{eqnarray}
Finally $\Omega_V$ is defined as:
\begin{equation} \label{eq:defmv}
 {\cal M}_{V \rightarrow \pi^+ \pi^- \pi^0} = i \varepsilon_{\mu \nu \alpha \beta} \,  p_1^\mu \, p_2^{\nu} \, p_3^\alpha \, \varepsilon_V^\beta \, \Omega_V \, ,
\end{equation}
being $\varepsilon_V^{\mu}$ the polarization of the vector meson.
Within our resonance chiral theory framework the corresponding reduced amplitudes, $\Omega_V$, are:
\begin{eqnarray} \label{eq:momega}
\Omega_{\omega}&=& \left(\sqrt{\frac{2}{3}}\cos \theta _V +\sqrt{\frac{1}{3}}\sin\theta _V \right)  \frac{8 \cos \delta}{M_{\omega} F^3}
\left\{\frac{\sqrt{2}}{M_V} G_{R\pi}(M_{\omega}^2) \right. \nonumber\\[3mm]
&&  + G_V \cos 2 \delta \,
BW[\rho,s]~D_{R\pi }(M_{\omega }^2,s) +G_V
BW[\rho,t]~D_{R\pi}(M_{\omega }^2,t) \nonumber \\[3mm]
&& + G_V
BW[\rho,u]~D_{R\pi }(M_{\omega }^2,u)  +2 G_V \sin^2 \delta \,
BW[\omega,s]~D_{R\pi}(M_{\omega }^2,s) \bigg\} ,\\[10mm]
\Omega_{\phi}&=&\left(\sqrt{\frac{1}{3}}\cos \theta _V-\sqrt{\frac{2}{3}}\sin \theta _V\right)  \frac{8}{M_{\phi} F^3} \left\{\frac{\sqrt{2}}{M_V}G_{R\pi}(M_{\phi }^2) \right.
\nonumber \\[3mm]
&&+ G_V  \cos^2 \delta \,
BW[\rho,s]~D_{R\pi }(M_{\phi }^2,s)
+ G_V BW[\rho,t]~D_{R\pi }(M_{\phi }^2,t) \nonumber \\[3mm]
&&+ G_V
BW[\rho,u]~D_{R\pi }(M_{\phi }^2,u)  + G_V\sin^2 \delta \,
BW[\omega,s]~D_{R\pi }(M_{\phi }^2,s) \bigg\},\\[10mm]
\Omega_{\rho}&=&\left(\sqrt{\frac{2}{3}}\cos \theta _V+\sqrt{\frac{1}{3}}\sin \theta _V \right)
\frac{8\sin \delta}{M_\rho F^3}\Bigg\lbrace \frac{\sqrt{2}}{M_V} G_{R\pi}(M_\rho^2) \nonumber \\[3mm]
&&+2 G_V \cos^2 \delta \,  BW_R[\rho,s]D_{R\pi}(M_\rho^2, s)+G_V BW_R[\rho,t]D_{R\pi}(M_\rho^2, t) \nonumber \\[3mm]
&&+G_V BW_R[\rho,u]D_{R\pi}(M_\rho^2, u)- G_V \cos 2\delta \,  BW_R[\omega,s]D_{R\pi}(M_\rho^2, s) \Bigg\rbrace ,
\end{eqnarray}
where $u=M_V^2+3 m_{\pi}^2-s-t$.



\end{document}